\documentclass[twocolumn,pre,showpacs,preprintnumbers,amsmath,amssymb]{revtex4}
\usepackage[xdvi]{graphicx}%
\usepackage{dcolumn}
\usepackage{amsmath}
\newcommand{\bra}{\left\langle}
\newcommand{\ket}{\right\rangle}

\newcommand{\pder}[2]{\frac{\partial #1}{\partial  #2}}
\newcommand{\pdert}[2]{\frac{\partial^2 #1}{\partial  #2^2}}

\newcommand{\dert}[2]{\frac{d^2 #1}{d  #2^2}}
\newcommand{\var}[2]{\frac{\delta #1}{\delta  #2}}

\newcommand{\e}{{\rm e}}

\newcommand{\phis}{\phi_*}
\newcommand{\phiB}{\phi_{\rm B}}
\newcommand{\phiBB}{\phi_{{\rm B}0}}
\newcommand{\ep}{\epsilon}

\newcommand{\tauw}{\tau_{\rm w }}

\newcommand{\Dt}{\Delta t}
\newcommand{\lr}[1]{\left(#1\right)}
\newcommand{\nm}{\nonumber\\}
\newcommand{\calF}{{\cal F}}

\newcommand{\lambdam}{\lambda_{\rm m}}

\newcommand{\spphi}{\phi_{\rm sp }}
\newcommand{\Fu}{F_{\rm u}}
\newcommand{\Vu}{V_{\rm u}}
\newcommand{\Wu}{W_{\rm u}}

\begin{document}

\title{Theoretical analysis for critical fluctuations 
of relaxation trajectory near a saddle-node bifurcation}

\author{Mami Iwata}
\email{iwata@jiro.c.u-tokyo.ac.jp}
\author{Shin-ichi Sasa}
\email{sasa@jiro.c.u-tokyo.ac.jp}
\affiliation
{Department of Pure and Applied Sciences,
University of Tokyo, 3-8-1 Komaba Meguro-ku, Tokyo 153-8902, Japan}
\date{\today}


\begin{abstract}
A Langevin equation whose deterministic part undergoes 
a saddle-node bifurcation
is investigated theoretically. 
It is found that statistical
properties of relaxation trajectories in this system
exhibit divergent behaviors 
near a saddle-node bifurcation point in the weak-noise limit, 
while the 
final value of the deterministic solution
changes discontinuously at the point.
A systematic formulation for analyzing a path 
probability measure is constructed
on the basis of a singular perturbation 
method. In this formulation, the critical nature turns out to 
originate from the neutrality of exiting time from a saddle-point.
The theoretical calculation explains results of
numerical simulations.
\end{abstract}

\maketitle

\section{introduction}

To uncover the nature of fluctuations near a bifurcation 
point has provided a clue to understanding of singularities 
 observed in a rich variety of phenomena. 
The most typical example of such studies
is the Ginzburg-Landau theory 
for equilibrium critical phenomena \cite{GL}. According to this 
theory, the description of fluctuations in the system that undergoes
a pitchfork bifurcation is a starting point for characterizing 
the Ising universality class of the paramagnetic-ferromagnetic transition. 
The second example is a theory of 
collective synchronization in coupled oscillators \cite{kuramoto}. 
In this phenomenon, the mechanism of the cooperative 
oscillation is explored by studying fluctuations near a 
Hopf bifurcation.  The third example is related to 
a theory of directed percolation, whose universality
class is characterized by a transcritical bifurcation 
with a multiplicative noise \cite{Munos}. 
From a viewpoint of bifurcation theory, a pitchfork bifurcation, 
a Hopf bifurcation, and a transcritical bifurcation  are local
 co-dimension one  bifurcations \cite{Gucken}. Now, the last 
one in this type of bifurcations is a saddle-node bifurcation.


Mathematically, a saddle-node bifurcation  in a differential equation 
is defined as the 
appearance of a pair of saddle type  fixed point  and   node type fixed 
point with respect to change in a system parameter. 
This bifurcation has been found in many models such as a mean field 
model of the spinodal transition \cite{binder}, a model of driven 
colloidal particles \cite{reimann}, bio-chemical network models 
\cite{bio3, spike, Tyson, Ohta_co},
a dynamical model associated with a
$k$-core percolation problem \cite{kcore}, and a 
random-field Ising model \cite{rfohta}. Theoretically, once it is 
found that a system undergoes a saddle-node bifurcation, its 
deterministic behavior  near the bifurcation point is immediately derived, 
as seen in standard textbooks \cite{Gucken}. As an example, 
a characteristic time scale exhibits divergent behavior 
proportional to  $\ep^{-1/2}$, where $\ep$ represents the distance 
from the bifurcation point.


Now, in a manner similar to the other local co-dimension one bifurcations,
it is expected that fluctuations near a saddle-node bifurcation exhibits 
a singular behavior. Thus far, the singular nature of the fluctuations has 
not been focused on except for some works
\cite{1dimlett,Ohta_co,kcore,rfohta,reimann,kubokitahara,spike}, 
and its theoretical 
study is still immature compared with much progress in 
theories of fluctuations near the other 
bifurcations. However, as we already pointed out
\cite{1dimlett,Ohta_co,kcore,rfohta}, a stochastic model 
under a saddle-node bifurcation is regarded as 
the simplest one of systems 
that exhibit the coexistence of discontinuous transition and critical fluctuation.
Such a coexistence has been observed in the 
dynamical heterogeneity in glassy systems 
\cite{chi4_allstar1,chi4_allstar2,silbert_mix,Toninelli_b, toninelli,schwartz,sellitto}.
Therefore, a theoretical method for describing fluctuations
near a saddle-node bifurcation might be useful 
for studying wider systems including glassy systems
(see Sec. \ref{discussion} as a related discussion).

With this background, in the present paper, we analyze a Langevin equation
for a quantity $\phi$ in which the deterministic
part undergoes a saddle-node bifurcation. Especially, 
we investigate statistical properties of relaxation behavior near the 
bifurcation point with small noise intensity $T$.
Note that $T$ is proportional to the inverse of the number of elements 
in  cases of many-body systems 
with an infinite range interaction \cite{Ohta_co} or 
defined on a random graph \cite{kcore,rfohta}.
Then, 
the fluctuation intensity $\chi_{\rm \phi}(t)$ of relaxation trajectories
exhibits a divergent behavior 
similar to those observed in the dynamical heterogeneity.
The aim of this paper is to describe this divergent behavior theoretically.


The main idea in our theoretical analysis is to express
a single trajectory in terms of 
a singular part and the others.
Concretely, we focus on 
the exiting time $\theta$ from a saddle-point and find that 
$\theta$ exhibits the divergent behavior 
$\bra (\theta-\bra \theta \ket)^2 \ket \simeq T^{-2/3}$  
in the limit $T \to 0$ with $\ep=0$ fixed and 
$\bra (\theta-\bra \theta \ket)^2 \ket \simeq \ep^{-5/2}$ 
in the limit $\ep \to 0$  
with $T \ll 1 $ fixed. In addition to these divergent behaviors,  
we derive a statistical distribution of $\theta$, by which
$\chi_\phi(t)$ is calculated.
This idea does not only make the calculation
possible, but also provides us an insight for the nature of the
divergent fluctuations near a saddle-node bifurcation. 
That is, the most important  quantity that 
characterizes the fluctuations near a saddle-node bifurcation is the 
exiting time from the saddle-point.

This paper is organized as follows. 
In Sec. \ref{model}, we present a model, and display
numerical results on divergent behavior of $\chi_{\phi}(t)$. 
In Sec. \ref{simple_analysis}, on the basis of 
a simple phenomenological argument, we derive 
critical exponents characterizing the divergent behavior.
We also present the 
basic idea of our theory.
In Sec. \ref{analysis}, by employing a singular perturbation 
method, we construct a systematic perturbation theory
so as to determine the statistical properties of 
 the important quantity, $\theta$.
Section \ref{discussion} is devoted to concluding remarks.

\section{model}
\label{model}

\begin{figure}[htbp]
\includegraphics[width=7cm]{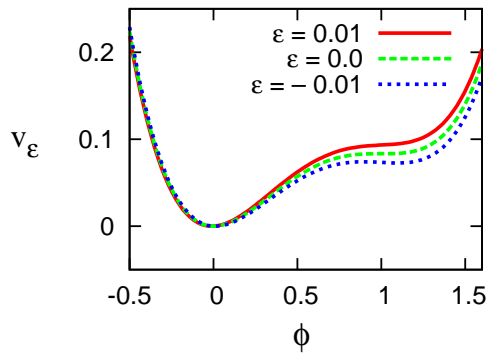}
\caption{
(color online).
Potential $v_\epsilon(\phi)$ for a few values of $\ep$.
}
\label{fig_pot}
\end{figure}

Let $\phi(t)$ be a time dependent one-component quantity. We study 
relaxation behavior described by a  Langevin equation 
\begin{eqnarray}
\partial_t\phi =f_\epsilon(\phi)+\xi.
\label{langevin}
\end{eqnarray}
Here, $f_\epsilon(\phi)$ is assumed to be 
\begin{eqnarray}
f_\epsilon(\phi)=-\epsilon \phi -\phi(\phi-1)^2,
\label{fformm}
\end{eqnarray} 
where $\ep$ is a small parameter.  $\xi$ in (\ref{langevin})
represents  Gaussian white noise that satisfies
\begin{eqnarray}
\bra\xi(t)\xi(t')\ket=2T\delta(t-t'),
\label{gauss_noise_model}
\end{eqnarray}
where $T$ represents the noise intensity.
A qualitative behavior of deterministic
trajectories for the system with 
$T=0$ is understood from the form of a potential 
defined by $f_\epsilon=-\partial_{\phi}v_\epsilon$, where 
the potential $v_\ep(\phi)$ is written as
\begin{equation}
v_\ep(\phi)=
\frac{\ep}{2}\phi^2+\frac{1}{2}\phi^2-\frac{2}{3}\phi^3+\frac{1}{4}\phi^4.
\end{equation}
As displayed in Fig.~\ref{fig_pot}, there is 
a unique  stable fixed point $\phi=0$ when $\ep > 0$, while 
there are two fixed points   $\phi=0$ and $\phi=1$ when $\epsilon=0$.
Since the latter is marginally stable, it is called a {\it marginal saddle}. 
Furthermore, when $\ep<0$, there are two stable fixed point
and one unstable fixed point.
This qualitative change of deterministic trajectories  at $\ep=0$ is 
called {\it saddle-node bifurcation}.
Throughout this paper, we assume the initial condition 
$\phi(0)=\phi_0 > 1$ and $\epsilon \geq 0$ so that the trajectories
pass the marginal saddle at $\phi=1$. 

We denote a trajectory  $\phi(t)$, $0 \le t \le \infty$, by $[\phi]$.
All the statistical quantities of trajectories are described by 
the path probability measure 
\begin{eqnarray}
 P([\phi]) = \frac{1}{Z_{\phi}}\e^{-\frac{\calF([\phi])}{T}},
\label{path_integral1}
\end{eqnarray}
where $Z_\phi$ is the normalization factor which is independent of $[\phi]$,
and ${\cal F}([\phi])$ is determined as
\begin{eqnarray}
\calF([\phi]) = \frac{1}{4}\int dt \lr{
\lr{(\partial_t \phi-f_\epsilon(\phi)}^2+2Tf_\ep'(\phi)}.
\label{path_integral2}
\end{eqnarray}
(See Appendix~\ref{path} for its derivation.) 
The last term in (\ref{path_integral2})
corresponds to the Jacobian associated with the transformation 
from a noise sequence $[\xi]$ to the corresponding
trajectory $[\phi]$.

\subsection{Numerical simulations}

Before entering the theoretical analysis of (\ref{langevin}), 
we report results of numerical simulations.
The Langevin equation was solved numerically with 
the Heun method \cite{heun}
with a time step $10^{-4}$,
and the initial condition was fixed as $\phi (0)=\phi_0=1.2$.
The expectation value $\bra A \ket$ of a fluctuating
quantity $A$ was
estimated as the average of one-hundred data. By using 
ten independent samples of the estimated values,
the value of $\bra A \ket$ is  conjectured  
with error-bars.

\begin{figure}[htbp]
\includegraphics[width=7cm]{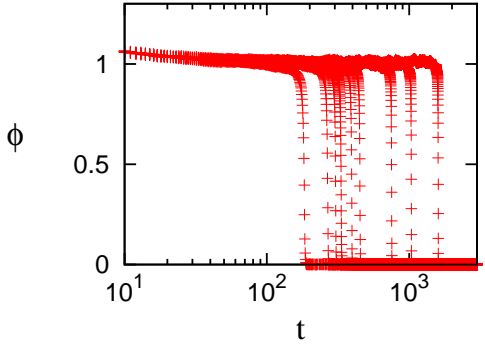}
\caption{(color online).
Trajectories $\phi(t)$ with $\ep=0$ and $T=2^{-20}$.
The nine trajectories are generated by
different noise sequences, respectively.
}
\label{1samp_path}
\end{figure}

In Fig.~\ref{1samp_path}, we display nine trajectories  
$\phi(t)$ for the system with $\ep=0$ and $T=10^{-6}$,
where each  trajectory is generated by a different noise sequence. 
The trajectories are clearly distinguished despite a rather small 
value of $T$. 
It is seen that a major difference among the trajectories is the exiting time 
from a region near $\phi=1$. Similar behaviors are 
observed  for other small values of $\ep$ and $T$.

\begin{figure}[htbp]
\includegraphics[width=7cm]{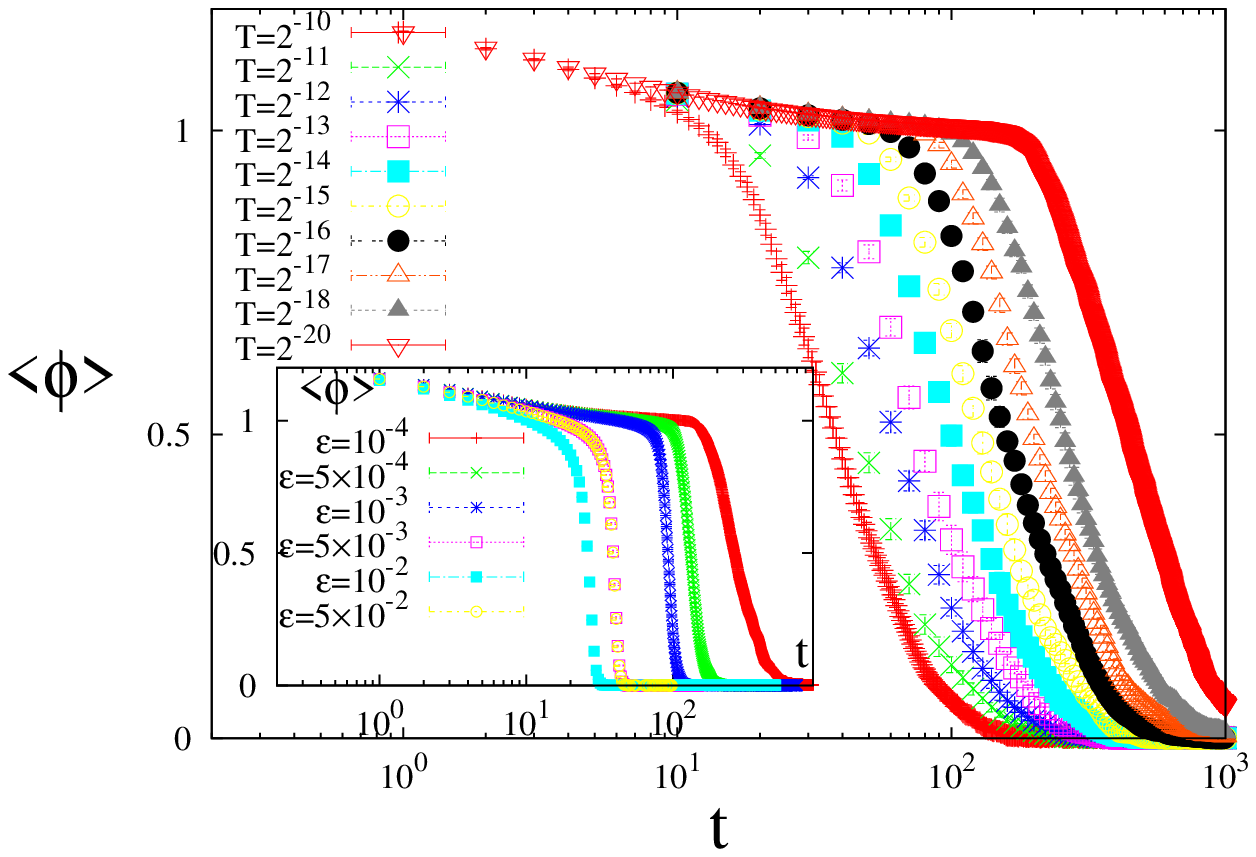}
\caption{(color online).
$\bra \phi(t) \ket$ for several values of $T$. $\ep=0$.
Inset: 
$\bra \phi(t) \ket $ for several values of $\ep$.
$T=2^{-20}$. 
}
\label{fig:phit0d}
\end{figure}

In order to clarify  $(\epsilon, T)$ dependence of the
relaxation behavior, we investigate $\bra \phi(t) \ket $
for several values of $\ep$ and $T$. As is seen from 
Fig.~\ref{fig:phit0d}, $\bra \phi(t) \ket $ exhibits 
the two steps relaxation,
and the plateau regime around $\phi\simeq 1$ 
becomes longer as $T$ is decreased with $\ep=0$ fixed 
(see Fig.~\ref{fig:phit0d}),
or as $\ep$ is decreased with $T=2^{-20}$ fixed
(see the inset of Fig.~\ref{fig:phit0d}).
These qualitative behaviors are easily conjectured 
from the form of the potential $v_{\rm \ep}(\phi)$.

\begin{figure}[htbp]
\includegraphics[width=7cm]{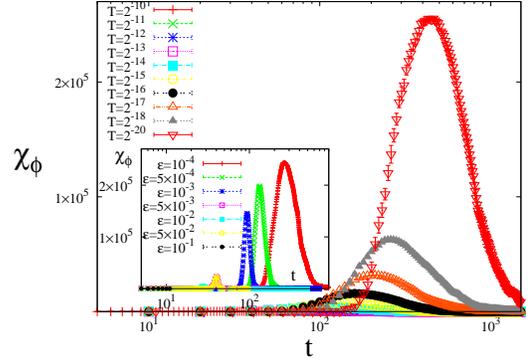}
\caption{(color online).
$\chi_{\phi} (t)$ for several values of $T$. $\ep=0$.
Inset: 
$\chi_{\phi} (t)$ for several values of $\ep$. $T=2^{-20}$.
}
\label{fig:chiphi0d}
\end{figure}

An impressive feature of the relaxation behavior is qualified
by the fluctuation intensity of $\phi(t)$:
\begin{eqnarray}
\chi_{\phi}(t) \equiv  T^{-1} (\bra\phi^2 (t)\ket - \bra\phi (t)\ket^2).
\label{eq:def_chiphi}
\end{eqnarray}
Note that $\chi_{\phi}(t)$ is independent of $T$ in the limit 
$T \to 0$ when fluctuations are not singular.
As displayed in Fig.~\ref{fig:chiphi0d},  
each $\chi_{\phi}(t) $ for small $\ep$ and $T$
takes a maximum value at a time $t_*$.
Furthermore, it is seen from these graphs that both the time 
$t_*$ and the amplitude $\chi_{\phi}(t_*)$ increase as 
$T$ is decreased with $\ep=0$ fixed 
(see Fig.~\ref{fig:chiphi0d}) or as $\ep$ is decreased 
with $T=2^{-20}$ fixed 
(see the inset of Fig.~\ref{fig:chiphi0d}).

%
%

\begin{figure}[htbp]
\includegraphics[width=7cm]{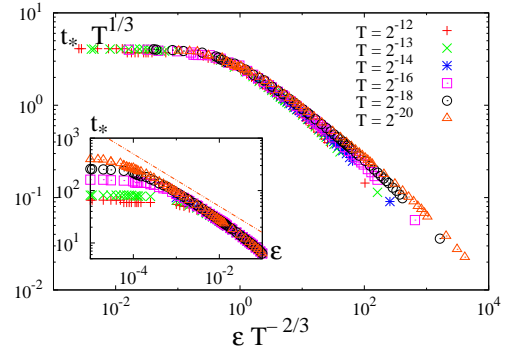}
\caption{(color online).
$t_* T^{1/3}$ as functions of $\ep T^{-2/3}$ for several values of $T$.
Inset: 
$t_*$ as functions of $\ep$.
The guide line represents a power-law function 
$t_* \simeq \ep^{-1/2}$.
}
\label{tau0}
\end{figure}

\begin{figure}[htbp]
\includegraphics[width=7cm]{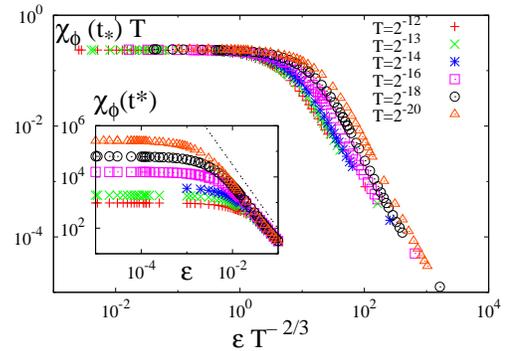}
\caption{(color online).
$\chi_{\phi}(t_*)T$ as functions of $\ep T^{-2/3}$ for several values of $T$.
Inset: 
$\chi_{\phi}(t_*)$ as functions of $\ep$.
The guide line represents a power-law function 
$\chi_{\rm \phi}(t_*)\simeq \ep^{-5/2}$.
}
\label{chi0}
\end{figure}

Now, we describe the divergent behaviors of 
$t_*$ and $\chi_\phi(t_*)$ quantitatively. We first note that 
the behaviors are classified into two regimes, 
(i) $\ep \ll T^{2/3} \ll 1$ and 
(ii) $T^{2/3} \ll \ep \ll 1$.
We observe 
\begin{eqnarray}
t_* &\simeq & T^{-1/3},
\label{ts:T}\\
\chi_{\phi}(t_*) & \simeq & T^{-1},
\label{chiphi:T}
\end{eqnarray}
in the regime (i), while 
\begin{eqnarray}
t_* &\simeq & \ep^{-1/2},
\label{ts:ep}\\
\chi_{\phi}(t_*) &\simeq & \ep^{-5/2},
\label{chiphi:ep}
\end{eqnarray}
in the regime (ii). The relations (\ref{ts:T}) and (\ref{chiphi:T})
are conjectured from the graphs in the mainframes of Figs.~\ref{tau0} 
and \ref{chi0}, respectively. 
Indeed, all the data of $t_* T^{1/3}$ and $\chi_{\rm \phi}(t_*)T$
seem to coincide with each other in the regime $\ep T^{-2/3} \ll 1$
for different small values of $T$.
The graphs in the insets of Figs.~\ref{tau0} and \ref{chi0} also
indicate the relations 
(\ref{ts:ep}) and (\ref{chiphi:ep}) in  the limit
$T\to 0$ with small $\ep$ fixed.

%
%

%
%

The specific purpose of  this paper is  to provide a theoretical 
understanding of the divergent behaviors given by  (\ref{ts:T}), 
(\ref{chiphi:T}), (\ref{ts:ep}), and (\ref{chiphi:ep}). Here,
we  note that $\bra \phi(t) \ket $ exhibits the discontinuous 
behavior from $\ep=0$ to $\ep \to 0+$ with $T=0$ fixed or from 
$T=0$ to $T \to 0+$ with $\ep=0$ fixed. 
The coexistence of the discontinuous  nature of 
$\bra\phi(t)\ket$ and the critical nature of $\chi_\phi(t_*)$ 
is a characteristic feature of stochastic dynamics
near a saddle-node bifurcation. 
Such a coexistence, which is called a {\it mixed order transition}, has been 
observed in several systems in the context of dynamical heterogeneity
\cite{chi4_allstar1,chi4_allstar2,silbert_mix,Toninelli_b,toninelli,
schwartz,sellitto}. 
See a related discussion in Sec.~\ref{discussion}.

\section{phenomenological analysis} \label{simple_analysis}

In this section, we present a phenomenological argument
for deriving the relations (\ref{ts:T}), 
(\ref{chiphi:T}), (\ref{ts:ep}), and  (\ref{chiphi:ep}). 
This argument also provides an essential idea 
behind a systematic perturbation method which
will be presented in Sec.~\ref{analysis}.


We first note that there are two small parameters $\ep$ and 
$T$ in this problem. Despite of this fact, 
the discontinuous nature of $\bra\phi(t)\ket$ causes
difficulties in the theoretical analysis. Indeed, we need a 
special idea so as to develop a perturbation method.
Let us recall the typical trajectories displayed 
in Fig.~\ref{1samp_path}. All the trajectories are kinked
in the time direction. Here, 
the kink position, which corresponds
to the exiting time from a region around $\phi=1$,
fluctuates more largely
than the other parts of the trajectories.
These observations lead to a natural idea that
a kink-like trajectory is first
identified as an unperturbed state. 

\begin{figure}[htbp]
\includegraphics[width=7cm]{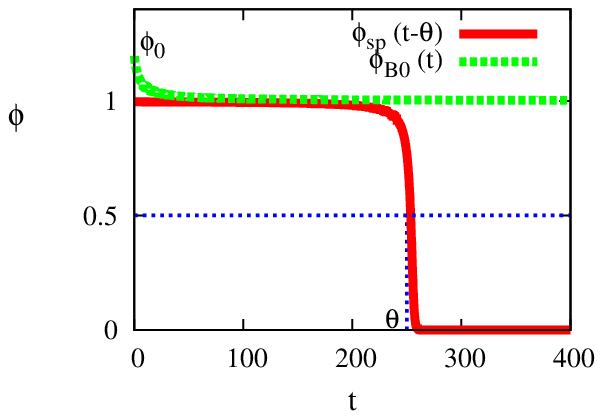}
\caption{(color online).
Functional forms of $\spphi (t-\theta)$ and $\phi_{\rm B0}(t)$. 
}
\label{phi_sp_phi_b}
\end{figure}

We illustrate the idea more concretely. 
Let us consider two special solutions of
$\partial_t\phi=f_0(\phi)$. The first 
is the special solution $\spphi(t)$ under the conditions 
$\spphi(t) \to 1$ as  $t \to -\infty$, $\spphi(t) \to 0$ as 
$t \to \infty$, and $\spphi(0)=1/2$. 
$\spphi(t) $
takes a kink-like form 
that connects the two fixed points. 
(See Fig.~\ref{phi_sp_phi_b}.)
The second one is the special solution $\phiBB(t)$ under the conditions 
$\phiBB(0) =\phi_0$ and $\phiBB(t) \to 1$ as 
$t \to \infty$.
$\phiBB(t) $  connects the initial value
and the marginal saddle. (See Fig.~\ref{phi_sp_phi_b}.) 
Then, by introducing a time $\theta$ which corresponds to a kink position, 
we express trajectories as 
\begin{equation}
\phi(t)=\spphi(t-\theta)+(\phiBB(t)-1)+\varphi(t-\theta),
\label{phit_assume_model}
\end{equation}
where $\varphi(t-\theta)$ represents deviation from the superposition
of the two special solutions. 


Now, fluctuations of $\phi(t)$ are expressed in terms of those of 
$\theta$ and $\varphi$. Then, it is reasonable to assume that 
$\varphi$ does not contribute largely to the divergent part 
of the fluctuation intensity of $\phi(t)$. With this assumption, 
$\bra \phi(t) \ket$ and $\chi_\phi(t)$ are estimated as
\begin{equation}
\bra \phi(t)\ket = \bra \spphi(t-\theta)\ket,
\label{avphi_assumtion1}
\end{equation}
and
\begin{equation}
\chi_\phi (t) =\frac{1}{T}\left(
\bra \spphi(t-\theta)^2\ket-\bra \spphi(t-\theta)\ket^2 \right), 
\label{avphi_assumtion2}
\end{equation}
where the statistical average is taken over $\theta$
with a distribution function of $\theta$. 
On the basis of these expressions,
we provide a phenomenological argument by which
we can determine the exponents characterizing the divergent behavior
of $t_*$ and $\chi_\phi(t_*)$.


The argument is divided into three parts. In Sec.~\ref{fluctuationsoftheta}, 
we first derive the scaling forms of $\bra \theta \ket $ and 
$\bra (\theta-\bra \theta \ket)^2 \ket$ in the two regimes, 
$T^{2/3} \ll \ep \ll 1$ and $\ep \ll T^{2/3} \ll 1$, respectively.
Second, based on these expressions,
we conjecture the distribution functions $P(\theta)$
for the two regimes. Finally, in Sec.~\ref{calculationofchiphi}, 
by using $P(\theta)$, we calculate the critical
exponents of $t_*$ and $\chi_\phi(t_*)$, which were observed numerically
in (\ref{ts:T}), (\ref{chiphi:T}), (\ref{ts:ep}), and (\ref{chiphi:ep}).

\subsection{Statistical properties of $\theta$} \label{fluctuationsoftheta}

The fluctuation intensity  of $\theta$ is defined as
\begin{equation}
\chi_\theta \equiv \frac{1}{T}(\bra \theta^2 \ket -\bra \theta \ket^2).
\label{chitheta_def}
\end{equation}
We then assume scaling relations 
\begin{eqnarray}
\bra \theta \ket  &=& T^{-\zeta'/\nu_*} f_1( T^{-1/\nu_*} \epsilon) , 
\label{sca1} \\
\chi_\theta  &=& T^{-\gamma'/\nu_*} f_2( T^{-1/\nu_*} \epsilon) ,
\label{sca2} 
\end{eqnarray}
for small $\ep$ and  $T$, with new exponents 
$\zeta'$, $\nu_*$, and $\gamma'$. Here, 
the scaling functions $f_1(x)$ and $f_2(x)$ satisfy two conditions:
(i) $f_1(0)$ and $f_2(0)$ are finite and
(ii) $f_1(x) \simeq x^{-\zeta'}$ and $f_2(x) \simeq x^{-\gamma'}$ 
for $x \gg 1$.
The latter condition implies that $\bra \theta\ket$ and $\chi_{\theta}$ 
are independent of T in the regime $T^{1/\nu_*} \ll \ep \ll 1$, where 
$\theta$ is expected to obey a Gaussian distribution with variance
proportional to $T$ (see (\ref{prob_thetas})).

\begin{figure}[htbp]
\includegraphics[width=7cm]{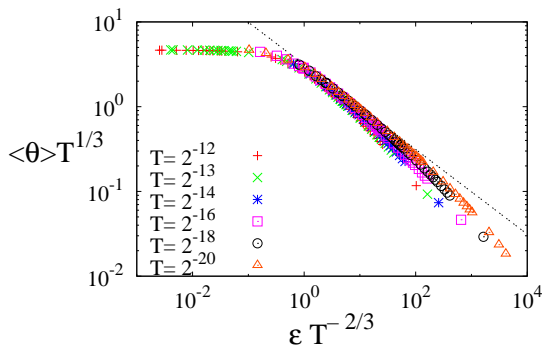}
\caption{(color online).
$\bra\theta\ket T^{1/3}$ as  functions of $\ep T^{-2/3}$. 
The guide line represents 
$\bra\theta\ket = 2 \lr{6^{1/4}}\ep^{-1/2}$, 
which is our theoretical result (\ref{dist_scale1}).
The graphs for six different values of $T$ tend to converge to one curve
as $T$ is decreased.
}
\label{theta_scale1}
\end{figure}
\begin{figure}[htbp]
\includegraphics[width=7cm]{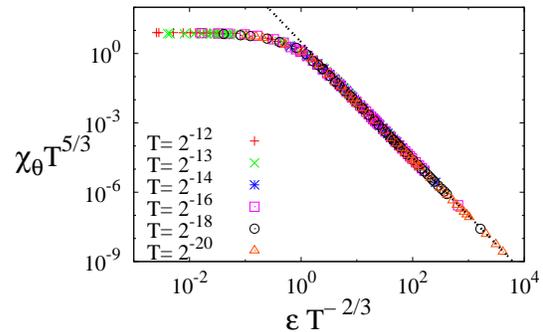}
\caption{(color online).
$\chi_{\rm \theta} T^{5/3}$ as functions of $\ep T^{-2/3}$.
The guide line represents 
$\chi_{\rm \theta}= 2 \lr{6^{1/4}}\ep^{-5/2}$, 
which is our theoretical result (\ref{dist_scale2}).
The graphs for six
different values of $T$ are on one curve. 
}
\label{theta_scale2}
\end{figure}

We shall determine the exponents $\zeta '$, $\nu_*$, and $\gamma '$
in (\ref{sca1}) and (\ref{sca2}).
Since a characteristic time around the marginal saddle is expected to 
obey the same scaling relation as $\theta$,
we investigate the local behavior near $\phi=1$.
Concretely, we substitute $\phi=1+\varphi$ into (\ref{path_integral2})
and ignore higher powers of $\varphi$
from an assumption that $|\varphi|$ is small.
We then obtain the probability 
measure for $[\varphi]$ as
\begin{eqnarray}
P\lr{[\varphi]}=\exp\lr{-{\tilde \calF}([\varphi])/T},
\end{eqnarray}
where
\begin{eqnarray}
\tilde \calF([\varphi]) = \frac{1}{4}\int dt \lr{
\lr{(\partial_t \varphi+\ep+\varphi^2}^2-4T \varphi}
+{\rm const}.
\label{path_integral3}
\end{eqnarray}
From this expression, when $T=0$,
we immediately find that $\varphi$ has a scaling form
$\varphi(t)= \ep^{1/2} \bar \varphi_1(\ep^{1/2}t)$.
Since the characteristic time scale in this case
diverges as $\ep^{-1/2}$,
we find that $\zeta'=1/2$. On the other hand, 
when $\ep=0$, we obtain another scaling form 
$\varphi(t)= T^{1/3} \bar \varphi_2(T^{1/3}t)$
which leads to $\zeta'/\nu_*=1/3$. We thus  derive
$\nu_*=3/2$. From this result, we expect that 
a distribution function of $\theta$ is expressed as 
a $T$-independent function of $\theta T^{1/3}$ 
when $\epsilon=0$. 
This expectation leads to a relation  
$\gamma' / \nu_* =2 \zeta' / \nu_* +1$, which 
yields $\gamma'=5/2$. These results are summarized as 
\begin{eqnarray}
\bra\theta \ket &=&  T^{-1/3} f_1 (\ep T^{-2/3}), 
\label{scaling_ep1} \\
\chi_\theta &=& T^{-5/3} f_2 (\ep T^{-2/3}),
\label{scaling_ep2}
\end{eqnarray}
which, in particular,
involve the results 
\begin{eqnarray}
\bra \theta \ket &\simeq&  T^{-1/3},
\label{scaling_T3} \\
\chi_{\theta} &\simeq & T^{-5/3},
\label{scaling_T4}
\end{eqnarray}
in the regime  $\ep \ll T^{2/3} \ll 1$,
while 
\begin{eqnarray}
\bra \theta \ket &\simeq&  \ep^{-1/2},
\label{scaling_ep3} \\
\chi_{\theta} &\simeq & \ep^{-5/2},
\label{scaling_ep4}
\end{eqnarray}
in the regime  $T^{2/3} \ll \ep \ll 1 $.
As shown in
Figs.~\ref{theta_scale1} and \ref{theta_scale2},
numerical results are consistent with 
(\ref{scaling_ep1}) and (\ref{scaling_ep2}).


Based on these  results (\ref{scaling_T3}), (\ref{scaling_T4}), 
(\ref{scaling_ep3}), and  (\ref{scaling_ep4}), we conjecture
functional forms of $P(\theta)$ 
from which we will calculate $\chi_\phi(t)$.


First, we consider the regime $\ep \ll T^{2/3} \ll  1$.
In order to simplify the argument, we focus on the case
that $\ep =0$ and $T \ll 1$, as the
representative
in the regime $\ep \ll T^{2/3}\ll  1$.
In this case, an exiting event from a region near $\phi=1$
occurs by effects of small noise.
We then naturally expect that trajectories reach to 
$\phi=1/2$ randomly as a Poisson process except for 
a short time regime in which the behavior depends
on initial conditions. That is, we conjecture that 
$\theta$ obeys the Poisson distribution when 
$\theta$ is much larger than some cut-off value 
$\theta_{\rm cut}$. Furthermore, from 
(\ref{scaling_T3}) and  (\ref{scaling_T4}),
the distribution function $P(\theta)$ is expressed
as  $P(\theta)={\tilde P} (\theta T^{1/3})$.
These considerations lead to
\begin{equation}
P(\theta)=\frac{T^{1/3}\e^{-a\theta T^{1/3}}}{Z_{\theta}^{(1)}}
\label{theta_dist_assume1}
\end{equation}
for the regime $\theta >\theta_{\rm cut}$. 
The positive constant $a$ and the normalization constant 
$Z_\theta^{(1)}$ 
are independent of $T$. We also assume that $\theta_{\rm cut}=O(T^{-1/3})$, 
which means $\int^{\infty}_{\theta_{\rm cut}}
d\theta P(\theta) \not =  1 $ in the limit $T \to 0$. We thus introduce
a $T$-independent parameter as
$\tilde \theta_{\rm cut}=\theta_{\rm cut} T^{1/3}$.
The expression (\ref{theta_dist_assume1}) will be derived
in Sec.~\ref{analysis}.


Next, we consider the regime $T^{2/3} \ll \ep \ll 1$.
When $\ep \neq 0$ and $T=0$, $P(\theta)$ 
is the delta function $\delta(\theta-\bra \theta\ket)$, because
there are no fluctuations. For sufficiently small $T$, the distribution
function is smeared slightly, and this leads to a conjecture
that $\theta$  obeys a Gaussian distribution 
\begin{eqnarray}
P(\theta)=\frac{1}{Z_{\theta}^{(2)}}
\e^{-\frac{1}{T}
 \frac{\lr{\theta-\bra \theta\ket}^2}{2\chi_\theta}},
\label{prob_thetas}
\end{eqnarray}
where $Z_\theta^{(2)}$ is a normalization constant.
The expression (\ref{prob_thetas}) will be derived in Sec.~\ref{analysis}.

\subsection{Calculation of $\chi_\phi(t)$ }
\label{calculationofchiphi}
On the basis of the results (\ref{theta_dist_assume1}) and (\ref{prob_thetas}),
we calculate the critical exponents of $t_*$ and $\chi_\phi(t_*)$.
Here, for convenience of calculation, we introduce
the Fourier transform of $\spphi(t)$ as 
\begin{eqnarray}
\spphi(t)=\int \frac{d\omega}{2 \pi} \tilde{\phi}_{\rm sp}(\omega)
\e^{i \omega t}.
\label{zexp}
\end{eqnarray}
Then, from  the assumptions (\ref{avphi_assumtion1}) and (\ref{avphi_assumtion2}), 
we estimate 
\begin{eqnarray}
\bra \phi(t) \ket
&=&
\int \frac{d\omega}{2 \pi}
\tilde{\phi}_{\rm sp}(\omega) \e^{i\omega t} 
\bra \e^{-i \omega \theta }\ket ,
\label{form1:1}
\end{eqnarray}
and 
\begin{eqnarray}
\bra \phi(t)^2 \ket
&=&  
\int \frac{d\omega}{2 \pi}
\int \frac{d\omega'}{2 \pi}
\tilde{\phi}_{\rm sp}(\omega) \tilde{\phi}_{\rm sp}(\omega')
\e^{i(\omega+\omega')t} \nm
&&
\bra \e^{-i(\omega+\omega') \theta }\ket .
\label{form2:1}
\end{eqnarray}
By substituting (\ref{form1:1}) and (\ref{form2:1}) into
(\ref{eq:def_chiphi}), we obtain
\begin{eqnarray}
\chi_{\phi}(t)&=&
\frac{1}{T}
\int \frac{d\omega}{2 \pi}
\int \frac{d\omega'}{2 \pi}
\tilde{\phi}_{\rm sp}(\omega) \tilde{\phi}_{\rm sp}(\omega')
\e^{i(\omega+\omega')t} \nm
&&\lr{
\bra \e^{-i(\omega+\omega') \theta }\ket 
-\bra \e^{-i\omega \theta }\ket \bra \e^{-i\omega' \theta }\ket 
}.
\label{form2:2}
\end{eqnarray}
In the following paragraphs, 
we shall calculate $\chi_\phi (t)$ for the two regimes 
$T^{2/3} \ll \ep \ll 1$ and $\ep \ll T^{2/3} \ll 1$,
respectively.


First, we consider the regime $\ep\ll T^{2/3} \ll 1$ with
setting $\ep =0$.
By using the distribution function (\ref{theta_dist_assume1}),
we have 
\begin{eqnarray}
\bra \e^{-i \omega \theta }\ket
&=&
\int_0^{\theta_{\rm cut}} d\theta e^{-i\omega\theta} P(\theta) 
\nm
&+&
\int_{\theta_{\rm cut}}^{\infty} d\theta 
\frac{T^{1/3}\e^{-a\theta T^{1/3}-i\omega\theta}}{Z_{\theta}^{(1)}}.
\label{theta_dist_assume3}
\end{eqnarray}
We assume that the first term of (\ref{theta_dist_assume3}) 
is neglected when $T \ll 1$.
Then, by performing the integration, we calculate 
\begin{eqnarray}
\bra \e^{-i \omega \theta }\ket
&=&
\frac{T^{1/3}}{Z_{\theta}^{(1)}}
\left[
\frac{e^{-
\theta_{\rm cut} T^{1/3}
(a+i\omega T^{-1/3})
}
}{T^{1/3}\lr{a+iT^{-1/3}\omega}}
\right].
\label{theta_poisson1}
\end{eqnarray}
Similarly, we obtain 
\begin{eqnarray}
\bra \e^{-i(\omega+\omega') \theta }\ket 
&=&
\frac{T^{1/3}}{Z_{\theta}^{(1)}}
\left[
\frac{e^{-
\theta_{\rm cut} T^{1/3}
(a+i\lr{\omega+\omega'} T^{-1/3})
}
}{T^{1/3}\lr{a+iT^{-1/3}\lr{\omega+\omega'}}}
\right].\nm
\label{theta_poisson2}
\end{eqnarray}
Furthermore,
since $\spphi (t)$ shows a quick change from $\spphi(t)=1$
to $\spphi(t)=0$ around the kink position $t=\theta$, we approximate 
$\spphi (t)$ as $\Theta(-t)$, where $\Theta(t)$ is the
Heaviside step function. With this approximation,
the Fourier transform of $\spphi (t)$ becomes
\begin{eqnarray}
{\tilde{\phi}_{\rm sp} }(\omega) = \lim_{b\to 0+} \frac{-1}{i\omega - b}.
\label{tildesp}
\end{eqnarray}
Here,
we introduce a scaled time ${\tilde t} = t T^{1/3}$.
We then substitute (\ref{theta_poisson1}), 
(\ref{theta_poisson2}), and (\ref{tildesp})
into (\ref{form2:2}).
The result is 
\begin{eqnarray}
&&\chi_\phi(\tilde t  T^{-1/3}) \nm
&= &
\frac{1}{T}
\int \frac{d\omega}{2 \pi}
\int \frac{d\omega'}{2 \pi}
\lim_{b\to 0+}\frac{1}{i\omega-b}\frac{1}{i\omega'-b}
\e^{i(\omega+\omega')T^{-1/3}(\tilde t-\tilde \theta_{\rm cut})} \nm
&&
\left(
\frac{1}{ Z_{\theta}^{(1)}}
\left[
\frac{e^{-
\tilde \theta_{\rm cut} a
}
}{\lr{a+iT^{-1/3}\lr{\omega+\omega'}}}
\right]\right.
\nm
&&
-\left.
\frac{1}{ \lr{Z_{\theta}^{(1)}}^2}
\left[
\frac{e^{-
2\tilde  \theta_{\rm cut} a}}
{\lr{a+iT^{-1/3}\omega}
\lr{a+iT^{-1/3}\omega'}}
\right]\right).
\label{chi_phi_cal2}
\end{eqnarray}
By the transformation of integrable variables,
${\tilde \omega} = \omega T^{-1/3}$ and ${\tilde \omega'} = \omega' T^{-1/3}$,
$\chi_{\phi}(t)$ is expressed as a scaling form 
$\chi_{\phi}(t) =\tilde \chi_{\phi}(T^{1/3} t-\tilde \theta_{\rm cut})/T$.
This implies that  $t_* =O(T^{-1/3})$ and 
$\chi_{\phi}(t_*)=O(T^{-1})$ in the limit $T \to 0$.
Furthermore, we expect that these exponents are valid
for  $\ep$ and $T$ that satisfy $\ep \ll T^{2/3} \ll 1$.
In this manner, we have obtained the results
consistent with (\ref{ts:T}) and (\ref{chiphi:T}).

We next consider the regime $T^{2/3} \ll \ep \ll 1$
by using the distribution function (\ref{prob_thetas}).
The Gaussian integration with respect to $\theta$ leads
to
\begin{eqnarray}
\bra \e^{-i\omega \theta }\ket&=&
\e^{-i \omega \bra \theta\ket} 
\e^{ -\frac{\chi_\theta \omega^2 T}{2}},
\label{theta_gauss1}
\end{eqnarray}
and
\begin{eqnarray}
\bra \e^{-i(\omega+\omega') \theta }\ket&=&
\e^{-i(\omega+\omega') \bra \theta\ket}
\e^{ -\frac{\chi_\theta(\omega+\omega')^2 T}{2}}.
\label{theta_gauss2}
\end{eqnarray}
Then, by substituting (\ref{theta_gauss1}) and (\ref{theta_gauss2}) 
into (\ref{form1:1}) and (\ref{form2:2}), we obtain
\begin{equation}
\bra \phi(t) \ket
=
\int \frac{d\omega}{2 \pi}
\tilde{\phi}_{\rm sp}(\omega)
\e^{i \omega (t-\bra \theta\ket)} 
\e^{ -\frac{\chi_\theta \omega^2 T}{2}},
\label{ave_phi}
\end{equation}
and 
\begin{eqnarray}
\chi_{\phi} (t)
&=&
\frac{1}{T}
\int \frac{d\omega}{2 \pi}
\int \frac{d\omega'}{2 \pi}
\tilde{\phi}_{\rm sp}(\omega)\tilde{\phi}_{\rm sp}(\omega')
\e^{i(\omega+\omega') (t-\bra \theta\ket)}\nm
&&
\lr{
\e^{ -\frac{\chi_\theta(\omega+\omega')^2 T}{2}}
-
\e^{ -\frac{\chi_\theta (\omega^2+\omega'^2) T}{2}}
}.
\label{form2:3}
\end{eqnarray}
From these, we derive 
\begin{eqnarray}
\chi_\phi(t)
&= & 
\chi_\theta \sum_{k=1}^\infty
\frac{1}{k!} 
\left( 
\left( \sqrt{\chi_\theta T} \right)^{k-1}
\partial_t^k \bra \phi(t) \ket \right)^2 .
\label{form2:3-2}
\end{eqnarray}
Here, let $\tauw=\sqrt{\chi_\theta T}$ be the width 
of the distribution of $\theta$. 
When $\tauw \ll 1$, the expressions
(\ref{ave_phi}) and (\ref{form2:3-2}) are further
simplified.
Indeed, we may estimate
\begin{equation}
\bra \phi(t) \ket = \spphi(t-\bra \theta \ket),
\label{phi_gauss_ok2}
\end{equation}
and
\begin{equation}
\chi_\phi(t) = \chi_\theta  (\partial_t \spphi(t-\bra\theta\ket) )^2.
\label{chi_phi_gauss1}
\end{equation}
These results show that $\chi_\phi(t)$ takes a maximum at 
$t=t_*$, where $t_* \simeq \bra \theta \ket \simeq \ep^{-1/2}$.
Furthermore, we obtain
$\chi_\phi(t_*) \simeq \chi_\theta  \simeq \ep^{-5/2}$.
These results are consistent with (\ref{ts:ep}) and (\ref{chiphi:ep}).
Note that they are valid in the regime $\tauw \ll 1$
which 
means $ T^{2/5} \ll \ep \ll 1$.
It seems that there is  no power-law behavior
in the regime $T^{2/3} \ll \ep \ll T^{2/5}$. 
In fact, Fig.~\ref{chi0} suggests that $\chi_{\phi} (t_*)$ 
does not converge to one universal curve in the whole
region. 

To this point, we have explained that 
the singular behavior of $\chi_\phi(t)$ is determined
by the statistical distribution of $\theta$.
Our analysis shows that 
$\theta$ is the most important quantity for characterization
of the divergent fluctuations near the saddle-node bifurcation.
This claim is also conjectured
from a fact that the statistical properties of 
$\theta$ are simpler than those of $\phi$. 
(Compare Figs.~\ref{tau0} and \ref{chi0} with
Figs.~\ref{theta_scale1} and \ref{theta_scale2}.)
Thus, we have focused our 
theoretical analysis on the derivation of the statistical properties
of $\theta$. Note that the scaling relations (\ref{scaling_T3}) and 
(\ref{scaling_ep3})  were 
derived in Refs.~\cite{kubokitahara, binder}, 
and arguments closely related to (\ref{scaling_T4}) and (\ref{scaling_ep4})  
were also presented in Refs.~\cite{reimann,spike}. 
In particular, all the statistical properties of $\theta$ are described 
by the analysis of the backward Fokker-Planck equation \cite{Gardner}. 
However, in the previous approaches, 
a perturbative calculation with small $\epsilon$ and $T$ seems 
quite complicated.
In such cases, it would be almost impossible to analyze spatially 
extended systems. 
In order to improve the situation, in the next section, we develop 
a systematic perturbative calculation for the distribution 
function of $\theta$ within the path-integral formulation
(\ref{path_integral1}) with (\ref{path_integral2}).

\section{Analysis}\label{analysis}   

Our theory basically relies on the idea mentioned in 
Sec.~\ref{simple_analysis}. 
That is, we start with the expression (\ref{phit_assume_model})
and derive 
the distribution function of the exiting time $\theta$. 
Formally, the derivation might be done by performing the integration 
of $P([\phi])$ with respect to $\varphi$ with $\theta$ fixed.
However, as far as we attempt, it seems difficult to 
carry out this integration by a standard  path integral method. 
One difficulty originates from the existence of a transient 
region before passing the marginal saddle. This 
contribution is described by the interaction of $\phiBB$ and
$\theta$, and yields a non-trivial distribution of $\theta$ 
in the regime $\theta \ll \bra \theta \ket$. The other
difficulty arises in calculation of a perturbative expansion
around the solution $\spphi$. Since the solution approaches
the marginal saddle in the limit $t\to -\infty$, the stability 
of the solution 
is marginal. In such a case, a naive perturbation induces 
a singularity. We thus need to reformulate the perturbation problem. 

In order to overcome the two difficulties, in this paper,
we employ a method of fictitious stochastic  processes
\cite{fictitious}. Concretely, we introduce a variable $\phi(t,s)$ 
with a fictitious time $s$ and define a fictitious Langevin equation
whose $s$-stationary distribution function is equal to the path probability 
measure  given  in (\ref{path_integral1}).
The Langevin equation is written as
\begin{eqnarray}
\partial_s \phi (t,s)=-\var{\calF([\phi])}{\phi}+\sqrt{2T}\eta (t,s),
\label{langevin_s_1}
\end{eqnarray}
with 
\begin{eqnarray}
\bra \eta(t,s) \eta(t',s') \ket
=  \delta(t-t')\delta(s-s').
\label{langevin_noise}
\end{eqnarray}
By substituting (\ref{path_integral2}) into (\ref{langevin_s_1}),
we write explicitly
\begin{eqnarray}
\partial_s{\phi}&=&\frac{1}{2}\partial_t^2\phi- F_{\ep, T}(\phi)
+\sqrt{2T}\eta(t,s),
\label{langevin_s_2}
\end{eqnarray}
where 
\begin{eqnarray}
F_{\ep, T}(\phi)
&=&  \frac{f'_{\ep}(\phi)f_{\ep}(\phi)}{2}+\frac{T}{2}f_{\ep}''(\phi) 
\nonumber \\
&=& 
\frac{1}{2}\phi(3\phi-1)(\phi-1)^3
+J_T(\phi)
+G_\ep(\phi),
\label{langevin_s_fet}
\end{eqnarray}
with
\begin{eqnarray}
J_T(\phi)&=&-\frac{T}{2}(6\phi-4),
\label{langevin_s_fet2}
\nm
G_\ep(\phi)&=&
\frac{1}{2}\epsilon^2 \phi+\epsilon \phi(\phi-1)(2 \phi-1).
\label{langevin_s_fet3}
\end{eqnarray}
By interpreting
$t$ as a fictitious space coordinate,
we regard (\ref{langevin_s_2}) as a reaction-diffusion 
system with the boundary condition $\phi(0,s)=\phi_0$.
Particularly, since the system is bistable, one may
employ techniques treating kinks in such systems 
\cite{ohtakawasaki, eiohta}.
As a result, 
as will be shown in subsequent sections,
the interaction between $\phiBB$ and $\theta$ can be formulated 
as a perturbation and the problem arising from the marginal 
stability of $\spphi$ can be treated in a proper manner. 
We note that interesting noise effects in kink dynamics
in bistable systems were reported in Ref.~\cite{polymer}.

As  discussed in Sec.~\ref{model}, the qualitatively
different behaviors were observed depending on the
regime either $\ep \ll T^{2/3} \ll 1$ 
or $T^{2/3} \ll \ep \ll 1$. 
Correspondingly, the perturbation theory is developed for each regime.
Since the basic idea behind calculation details is in common
to both the regimes, 
we provide the full account of the perturbation
theory for the regime  $\ep  \ll T^{2/3} \ll 1$ in Sec.~\ref{formulation}, 
\ref{perturb}, and \ref{results}.
Then, we discuss briefly a perturbation theory
for the regime
$T^{2/3} \ll \ep \ll 1$ with pointing out  the difference from
the regime $\ep \ll T^{2/3} \ll 1$ in Sec. \ref{remark}.

\subsection{Formulation} \label{formulation}

\subsubsection{Unperturbed system} \label{unperturb}

\begin{figure}[htbp]
\includegraphics[width=7cm]{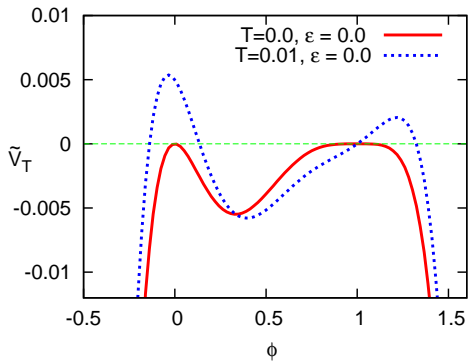}
\caption{(color online).
Potential $\tilde V_T(\phi)$ for $T=0$ and $T=0.01$.}
\label{fig_potv}
\end{figure}

\begin{figure}[htbp]
\includegraphics[width=7cm]{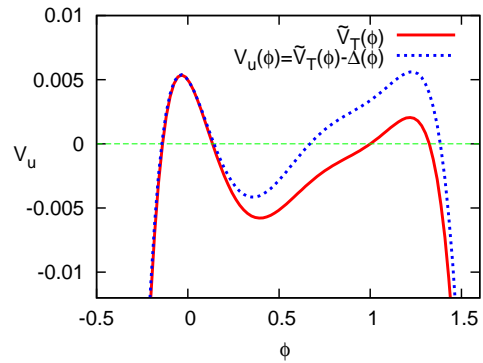}
\caption{(color online).
Unperturbative potential $\Vu(\phi)$, and 
${\tilde V}_T(\phi)$
for $T=0.01$.
}
\label{fig_potd}
\end{figure}

Since we focus on the regime $\ep \ll T^{2/3} \ll 1$, one may 
choose the system with $\epsilon=0$ and $T=0$ as an unperturbed 
system. However, we cannot develop a perturbation theory
with this choice. In order to explain the reason more explicitly,
we define a potential function $\tilde V_T(\phi)$ by
\begin{eqnarray}
F_{\ep=0, T}(\phi)= -\partial_{\phi}\tilde V_T(\phi), 
\label{ft_def}
\end{eqnarray}
where $\tilde V_T(1)=0$.
This potential is calculated as 
\begin{eqnarray}
\tilde V_T(\phi)=-\frac{1}{4}\phi^2(\phi-1)^4+\frac{T}{2}(\phi-1)(3\phi-1).
\label{vt_def}
\end{eqnarray}
In Fig.~\ref{fig_potv}, the functional forms of $\tilde V_T(\phi)$ 
are displayed  for $T=0$ and $T=0.01$. 
Here, the special solution $\spphi(t)$ in 
(\ref{phit_assume_model}) 
connects
the two maxima of $\tilde V_{T=0}(\phi)$. However, since the curvature 
of the potential $\tilde V_{T=0}(\phi)$  at the maximal point $\phi=1$ 
is zero, 
a perturbative correction to the solution $\phi_{\rm sp}(t)$ 
exhibits a divergent behavior. 
(See an argument 
below (\ref{green_explititform}).)
In order to avoid the singularity, 
 we must choose an unperturbed potential different from
$\tilde V_{T=0}(\phi)$.
Based on a fact that $\tilde V_T(\phi)$ 
has the two maxima at $\phi=\phi_1=-4T+O(T^2)$ 
and $\phi=\phi_2=1+T^{1/3}+O(T^{2/3})$, 
where $\tilde V_T(\phi_1)=T/2$ and 
$\tilde V_T(\phi_2)=3T^{4/3}/4+O(T^{5/3})$,
we define an unperturbed potential $\Vu$ by the  decomposition 
\begin{eqnarray}
\tilde V_T(\phi)=\Vu(\phi)+\Delta(\phi),
\label{decom}
\end{eqnarray}
where $\Delta(\phi)$ is chosen such that 
the two maximal values at $\phi=\phi_1$ and $\phi=\phi_2$ of
$V_u(\phi)$ 
are identical; that is, $\Vu(\phi_1)=\Vu(\phi_2)=T/2$
and $\Vu'(\phi_1)=\Vu'(\phi_2)=0$, as shown in Fig.~\ref{fig_potd}.
Furthermore, in order to have a simple argument, 
we impose a condition that the curvature at each maximum of $\Vu$ is 
equal to that of the corresponding maximum of $\tilde V_T$ 
with ignoring contribution of $O(T^{4/3})$. 
These conditions are satisfied  by setting 
$\Delta(\phi_1)=0$, $\Delta(\phi_2)= \tilde V_T(\phi_2)-\tilde V_T(\phi_1)
= 3 T^{4/3}/4-T/2+O(T^{5/3})$,  
$\Delta'(\phi_1)=\Delta'(\phi_2)=0$,
$\Delta''(\phi_1)=O(T^{4/3})$, and
$\Delta''(\phi_2)=O(T^{5/3})$.
As an example, one may choose 
\begin{eqnarray}
\Delta(\phi)=\left(\frac{3}{4}T^{4/3}-\frac{T}{2}+O(T^{5/3}) \right)
(12 v_{\ep=0}(\phi)+O(T)).
\label{delta}
\end{eqnarray}
By using the decomposition (\ref{decom}) and defining 
$\Fu\equiv -\partial_\phi \Vu$,  we rewrite 
(\ref{langevin_s_2}) as
\begin{eqnarray}
\partial_s\phi&=&\frac{1}{2}\partial_t^2{\phi}- \Fu(\phi)
-G_{\ep}(\phi)+\Delta' (\phi)+\sqrt{2T}\eta.
\label{langevin_s_3}
\end{eqnarray}
In our formulation, we treat  the last three  terms 
as perturbations. 

\subsubsection{Expression of solutions} 

\begin{figure}[htbp]
\includegraphics[width=7cm]{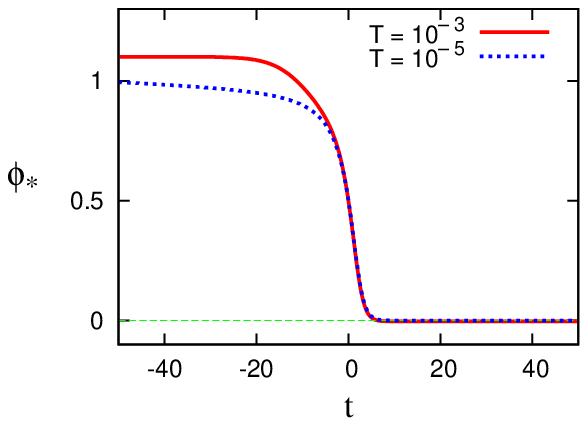}
\caption{(color online).
$\phis$ as  functions of $t$.
}
\label{phis_num}
\end{figure}

Since we have replaced the unperturbative potential, 
we reconsider an expression of trajectories.
We first define  the special  solution 
$\phis$ of the unperturbed equation 
\begin{eqnarray}
\frac{1}{2}\dert{\phis}{t}- \Fu(\phis)=0
\label{s-sol}
\end{eqnarray}
under the conditions $\phis(t) \to \phi_2$ when $t \to -\infty$
and $\phis(t) \to \phi_1$ when $t \to \infty$. 
We also impose $\phis(0)=1/2$ in order to determine $\phis$ uniquely. 
This solution corresponds to the kink solution in the real time direction
and describes the relaxation behavior from $\phi=\phi_2(\simeq 1)$ to 
$\phi =\phi_1(\simeq 0)$.  The functional form
of $\phis$ can be obtained 
by the integration of 
\begin{equation}
\partial_t\phis = -2\sqrt{\Vu(\phi_1)-\Vu(\phis)}.
\label{phi_def_app}
\end{equation}
In Fig.~\ref{phis_num}, we show $\phis (t)$ for $T=10^{-3}$ and $10^{-5}$.

Next, let $\phiB$ be the 
 solution of 
\begin{eqnarray}
\frac{1}{2}\pdert{\phiB}{t}- \Fu(\phiB)=0
\label{s-sol-B}
\end{eqnarray}
under the conditions that $\phiB(t) \to \phi_2$ as $t \to \infty$
and that $\phiB(0)=\phi_0$. 
This solution describes a typical 
behavior of $\phi$ near $t=0$ when $T$ is sufficiently small. 
By using these 
two solutions, we express the solution 
of (\ref{langevin_s_3}) for a given $\eta$ as 
\begin{eqnarray}
\phi(t,s) = \phis(t-\theta(s)) +(\phiB(t)-\phi_2)+
\rho(t-\theta(s),s),
\label{exp-sol}
\end{eqnarray}
where $\rho$ represents a possibly small deviation from 
the superposition of the two solutions. 

\subsubsection{Linear stability analysis} \label{linear stability analysis}

\begin{figure}[htbp]
\includegraphics[width=7cm]{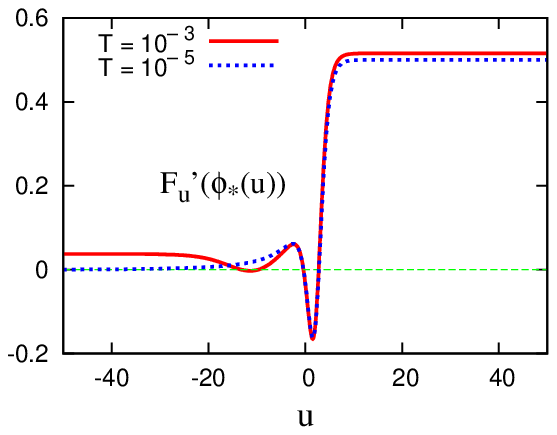}
\caption{(color online).
$F'_{\rm u}(\phis(u))$ as functions of $u$. 
}
\label{f0dphis}
\end{figure} 

As a preliminary for a systematic perturbation theory,
we perform the linear stability analysis of $\phis$.
Hereafter, we set $u=t-\theta$. The stability of the solution $\phis$ 
is determined by eigenvalues of the linear operator $\hat L$ given by
\begin{eqnarray}
\hat L \equiv \frac{1}{2}\frac{d^2}{du^2} - F'_u(\phis(u)).
\label{def_L}
\end{eqnarray}
Let us consider the eigenvalue problem 
\begin{eqnarray}
\hat L \Phi(u)=-\lambda\Phi(u).
\end{eqnarray}
This problem is equivalent to an energy eigenvalue problem 
in one-dimensional quantum mechanics, where 
$F_u'(\phis(u))$ and $\lambda$ correspond to 
the potential and an energy eigenvalue, respectively.  The graphs of 
$F_u'(\phis(u))$ are shown in Fig.~\ref{f0dphis}.
The asymptotic behaviors are calculated as 
$F_u'(\phis(u)) \to 1/2+O(T)$ as $ u \to \infty$,
and  $F_u'(\phis(u)) \to 3T^{2/3}+O(T)$ as $ u \to -\infty$.

\begin{figure}[htbp]
\includegraphics[width=7cm]{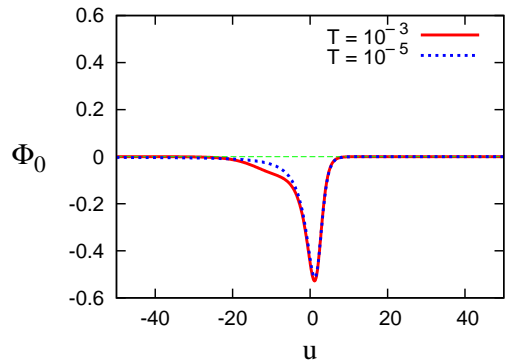}
\caption{(color online).
$\Phi_0$ as functions of $u$.
}
\label{dphis_num}
\end{figure}

Since $\phis(t-\theta)$ for any $\theta$ is 
a solution of (\ref{s-sol}), we may take the derivative 
of (\ref{s-sol}) with respect to $\theta$.
We then have
\begin{eqnarray}
\hat L \partial_u\phis=0.
\label{zero}
\end{eqnarray}
This implies that there exists the zero-eigenvalue,
for which the normalized eigenfunction $\Phi_0 (u)$ is determined as 
\begin{eqnarray}
\Phi_0 (u) &=&
\lr{\partial_u\phis}/\sqrt{\Gamma},
\label{zero_def}
\end{eqnarray}
where 
\begin{eqnarray}
\Gamma &= & \int_{-\infty}^\infty du (\partial_u\phis(u))^2.
\label{Gamma} 
\end{eqnarray}
Here, by using $\partial_u \phis=f_{\ep=0}(\phis)+O(T)$,
we obtain
\begin{eqnarray}
\Gamma &=&\int_{1}^{0} d\phi f_{\ep=0}(\phi)+O(T)\nm
&=&v_{\ep=0}(\phi)|_{0}^{1}+O(T)\nm
&=&\frac{1}{12}+O(T).
\end{eqnarray}
The zero-eigenfunction $\Phi_0 (u)$ corresponds to the Goldstone mode
associated with a time translational symmetry.
The functional form of $\Phi_0 (u)$ is shown in Fig.~\ref{dphis_num}.
Since there is no node in the $u$ profile of $\Phi_0 (u)$,
the minimum eigenvalue must be zero as in quantum mechanics.
Since the height of the potential in the limit $u\to -\infty$ is $O(T^{2/3})$,
it is expected that there is no other discrete eigenvalue  when $T$ is 
sufficiently small.
Next, we consider continuous eigenvalues for
$\lambda \ge \lambdam=\lim_{u\to\infty} F_u'(\phis(u)) = 3T^{2/3}+O(T)$. 
The corresponding eigenfunctions are characterized by 
the asymptotic plane waves with the eigenvalues 
$\lambda =\lambdam+k_{\rm L}^2/2$ and $\lambda =(1+k_{\rm R}^2)/2$,
where $k_{\rm R}$ and $k_{\rm L}$ are non-zero real numbers
representing wavenumbers of the asymptotic plane waves,
in the limit $u\to-\infty$ and $u\to+\infty$
respectively. 

We denote by $\Phi_\lambda(u)$ the eigenfunction corresponding to 
the eigenvalue $\lambda$. Since the eigenvalues greater than 
$1/2$ are degenerated,  $\Phi_\lambda^*(u)\not = \Phi_\lambda(u)$ 
for $\lambda \ge 1/2$, where ${}^*$ represents the complex conjugate.  
Here, it is convenient to introduce  $\Phi_{-\lambda}(u)$ as 
$\Phi_{-\lambda}(u)=\Phi_{\lambda}^*(u)$. Related to
this introduction, we also define the index set 
$\Lambda= \{\lambda | \lambda \le -1/2, \quad \lambda \ge \lambdam \}$.
Then, we may choose the set of eigenfunctions so as to satisfy the
orthogonality condition
\begin{eqnarray}
\int_{-\infty}^\infty 
du \Phi_{\lambda'}^*(u)\Phi_\lambda(u) = \delta(\lambda-\lambda'),
\end{eqnarray}
and
\begin{eqnarray}
\int_{-\infty}^\infty 
du \Phi_{0}(u)\Phi_\lambda(u) =0,
\end{eqnarray}
for any $\lambda,\lambda' \in \Lambda$. 
Furthermore, we expect the completeness condition
\begin{eqnarray}
\Phi_0(u)\Phi_0(u')
+\int_{\Lambda}d \lambda \Phi_\lambda^{*}(u)\Phi_\lambda(u')= \delta(u-u') .
\end{eqnarray}
By using these eigenfunctions, we expand $\rho(u,s)$ as 
\begin{eqnarray}
\rho(u,s)=\int_{\Lambda}d \lambda \psi_\lambda(s)\Phi_\lambda(u),
\label{rho-exp}
\end{eqnarray}
where the contribution of the zero eigenfunction 
is not taken into account in the expression (\ref{rho-exp})
so that the expression (\ref{exp-sol}) is uniquely determined.

\subsection{Perturbation theory}\label{perturb}     %


By substituting (\ref{exp-sol}) into (\ref{langevin_s_3}),
we obtain 
\begin{eqnarray}
&&
-\Omega \partial_u{\phis}+\partial_s\rho\nm
&=&
\hat L(\phis)\rho-\frac{1}{2}F''_{\rm u}(\phis)\rho^2  \nonumber \\
& &-G_{\ep}(\phis) +\Delta' (\phis) \nonumber  \\
& &+\hat L(\phis)(\phiB(t)-\phi_2) +\sqrt{2T}\eta+\cdots ,
\label{per:start}
\end{eqnarray}
where we set 
\begin{equation}
\Omega=\partial_s\theta.
\label{omegadef}
\end{equation}

We consider a perturbation  expansion of $\rho$ and $\Omega$
with respect to a small parameter. In the present problem, 
$G_{\ep}(\phis)$, $\Delta' (\phis)$,
$\hat L(\phis)(\phiB(t)-\phi_2)$, and $\sqrt{2T}\eta$ are 
treated as perturbations.
In order to formulate the problem concretely, we introduce a 
formal expansion parameter $\mu$ in front of $G_{\ep}(\phis)$, 
$\Delta' (\phis)$, $\hat L(\phis)(\phiB(t)-\phi_2)$,
and $T$ in the noise term. We solve the equation 
(\ref{per:start}) perturbatically under the assumption that
$\rho$ and $\Omega$ can be expanded in $\mu$.
Note that $\rho$=$\Omega$=0 when $\mu=0$.
Concretely, due to the small noise term 
which is $O(\mu^{1/2})$, we assume 
\begin{eqnarray}
\Omega &=&  \mu^{1/2} \Omega^{(1/2)}+\mu \Omega^{(1)}+\cdots, 
\label{exp-omega}\\
\rho &=&  \mu^{1/2} \rho^{(1/2)}+\mu \rho^{(1)}+\cdots 
\label{exp-rho}.
\end{eqnarray}
As we will see below, all the terms of $\rho^{(i)}$ and $\Omega^{(i)}$
can be calculated  in principle.  Here,
it should be noted that $\mu$ is not directly related to 
small parameters $\ep$ and $T$. Therefore, for example,
if one wishes
to derive the solution valid up to $O(T^{2})$,
we do not have a quick answer to the question how many
orders of expansion in $\mu$ are necessary. 
Although this aspect makes the analysis complicated, 
we will find that the formulation leads to a systematic
expansion.
These preliminaries presented in this section are standard in 
a singular perturbation method \cite{hohenberg,kuramoto_p}.
With this setting up, we calculate $\rho^{(i)}$ and $\Omega^{(i)}$ in sequence.

\subsubsection{Lowest order result} \label{lowest} %

We start with the calculation of $\Omega^{(1/2)}$ and $ \rho^{(1/2)}$.
We substitute the expansions (\ref{exp-omega}) and (\ref{exp-rho})
into (\ref{per:start}), arrange terms according to powers of $\mu$,
and pick up all the terms proportional to $O(\mu^{1/2})$.
We then obtain 
\begin{equation}
-\Omega^{(1/2)} \partial_u{\phis}+\partial_s\rho^{(1/2)}
=
\hat L(\phis)\rho^{(1/2)}+\sqrt{2T}\eta.
\label{eq:1-2}
\end{equation}
We rewrite (\ref{eq:1-2}) as a linear equation for $\rho^{(1/2)}$
with the expression
\begin{equation}
\lr{\partial_s-\hat L(\phis)}\rho^{(1/2)}
= B(\phis, \eta).
\label{eq:1-2b}
\end{equation}
Because $\hat L(\phis)$ possesses the zero-eigenvalue, 
it is not the case that there exists a unique bounded $\rho^{(1/2)}$. 
That is, there exists no bounded solution or
there are an infinitely number of solutions. In order to
proceed the calculation further by obtaining $\rho^{(1/2)}$,
we impose the solvability condition 
under which the latter case is chosen. 
The solvability condition in this case is written as 
\begin{equation}
\int_{-\infty}^\infty du \Phi_0(u) B(\phis,\eta)=0.
\label{rho1_2_linear_eq}
\end{equation}
This yields
\begin{equation}
- \Gamma \Omega^{(1/2)} =\sqrt{2T\Gamma}\eta_\theta,
\label{sol:1-2}
\end{equation}
where 
\begin{eqnarray}
\eta_\theta (s) &=& \int_{-\infty}^\infty
du (\partial_u \phis) \eta (u+\theta, s)
/\sqrt{\Gamma}.
\label{noise-theta}
\end{eqnarray}
We note that  (\ref{noise-theta}) satisfies 
\begin{eqnarray}
\bra \eta_\theta(s) \eta_\theta (s')   \ket = \delta(s-s').
\label{noise-theta-int}
\end{eqnarray}

Furthermore, under the solvability condition, one can determine statistical
properties of $\rho^{(1/2)}$ from (\ref{eq:1-2b}). In a manner similar to 
(\ref{rho-exp}), we expand $\rho^{(1/2)}$ in terms of 
the eigenfunctions of $\hat L (\phis)$. Then, for any non-zero eigenvalue
$\lambda$, the coefficient $\psi_\lambda^{(1/2)}$
obeys a Langevin equation
\begin{eqnarray}
\partial_s \psi_\lambda^{(1/2)}=-\lambda \psi_\lambda^{(1/2)}
+ \sqrt{2T}\eta_\lambda,
\label{psiveol}
\end{eqnarray}
with 
\begin{eqnarray}
\eta_\lambda (s)=\int_{-\infty}^\infty
du \Phi_{\lambda}(u)\eta(u+\theta,s).
\label{zetalambda_def}
\end{eqnarray}
From this, we obtain $\bra \psi_\lambda^{(1/2)} \ket_{\rm \theta} =0$ and 
\begin{eqnarray}
\bra  \psi_\lambda^{(1/2)}  \psi^{(1/2)}_{\lambda'}{}^* \ket_{\rm \theta}
= \frac{T}{\lambda}\delta(\lambda-\lambda'),
\end{eqnarray}
where $\bra \ket_{\rm \theta}$ denotes an expectation value with $\theta$
fixed. The fluctuation intensity of $\rho^{(1/2)}$ is then calculated as 
\begin{eqnarray}
\bra (\rho^{(1/2)})^2\ket_\theta &=&\int_{\Lambda}d\lambda
\frac{T}{\lambda}|\Phi_\lambda(u)|^2 \nm
&=&-T G(u,u),
\label{rho_1/2_green}
\end{eqnarray}
where $G(z,y)$ is the  Green function defined as
\begin{eqnarray}
G(z,y)=-\int_{\Lambda}d\lambda \frac{1}{\lambda}
\Phi_\lambda(z)\Phi_\lambda^*(y).
\label{Green}
\end{eqnarray}
Here, we note that $G(z,y)$ satisfies 
\begin{eqnarray}
\left(\frac{1}{2}\partial_z^2-\Fu'(\phis(z)) \right)
G(z,y) = \delta(z-y)-\Phi_0(z)\Phi_0(y).\nm
\end{eqnarray}
Let us estimate $G(u,u)$ in the limit $ u \to \pm \infty$
by defining 
\begin{eqnarray}
m_{\pm} = \lim_{u \to \pm \infty}\sqrt{2 F'_u(\phis(u))} .
\end{eqnarray}
From the expression of $\Fu(\phis(u))$ determined by 
 (\ref{langevin_s_fet}), (\ref{decom}), and (\ref{delta}), 
we calculate
\begin{eqnarray}
m_{+} &=&  1+O(T), \\
m_{-} &=&  \sqrt{6}T^{1/3}+ O(T^{2/3}),
\label{def_m_pm}
\end{eqnarray}
in the limit $T \to 0$. 
We then define 
a Green function $G_{\pm}(z,y)$ by
\begin{eqnarray}
\left(\frac{1}{2}\partial_z^2 - \frac{m_{\pm}^2}{2} \right)
G_{\pm} (z,y) = \delta(z-y).
\end{eqnarray}
This Green function is written as
\begin{eqnarray}
G_{\pm}(z,y)=-\frac{1}{m_{\pm}}\e^{-m_{\pm}|z-y|} .
\label{green_explititform}
\end{eqnarray}
We conjecture that $G(u,u)$ approaches $G_{\pm}(u,u)$
as $u\to \pm \infty$.
Then, from (\ref{rho_1/2_green}), we find
that $\bra (\rho^{(1/2)})^2\ket_\theta \simeq O(T^{2/3})$
for $u \to -\infty$, and $\bra (\rho^{(1/2)})^2\ket_\theta \simeq O(T)$
for $u \to \infty$. 

Here, we address one remark.
If $\spphi(t)$ were chosen as the unperturbative solution, $m_-$
would become zero, and therefore 
$\rho^{(1/2)}(u \simeq -\infty)$ 
would exhibit unbounded Brownian motion as a function of $s$.
This singularity originates from the marginal stability of 
$\spphi$. 
This is why we choose $\phi_*$ instead of $\spphi$
as the unperturbative solution, 
which was mentioned in Sec.~\ref{unperturb}.

\subsubsection{Next  order calculation} \label{The next  order cal} %

In the lowest order description, 
the variable $\theta$ exhibits unbounded Brownian motion,
and therefore it indicates a singular behavior.
Now, in order to determine the exponents characterizing the singularity,
we proceed to the next order calculation.
We substitute (\ref{exp-omega}) and (\ref{exp-rho}) into (\ref{per:start}) 
and pick up all the terms proportional to $O(\mu)$. We then obtain 
\begin{eqnarray}
-\Omega^{(1)} \partial_u{\phis}+\partial_s\rho^{(1)}
&=&
\hat L(\phis)\rho^{(1)}-\frac{1}{2}\Fu''(\phis)(\rho^{(1/2)})^2  \nonumber \\
& &-G_{\ep}(\phis) +\Delta' (\phis) \nm
& & +\hat L(\phis)(\phiB(t)-\phi_2).
\label{eq:1}
\end{eqnarray}
The solvability condition for the linear equation for 
$\rho^{(1)}$ yields 
\begin{eqnarray}
-\Gamma \Omega^{(1)}&=& \Psi_1+\Psi_2+\Psi_3+\Psi_4,
\label{sol:1}
\end{eqnarray}
where 
\begin{eqnarray}
\Psi_1 &=& -\frac{1}{2} \int_{-\infty}^\infty
du  (\partial_u \phis) F''_u(\phis)(\rho^{(1/2)})^2, 
\label{omega1} \\
\Psi_2 &=&  \int_{-\infty}^\infty du (\partial_u \phis) \Delta'(\phis), 
\label{omega2} \\
\Psi_3 &=&  \int_{-\infty}^\infty
 du (\partial_u \phis) \hat L(\phis)(\phiB(u+\theta)
-\phi_2),
\label{omega3} \\
\Psi_4 &=&  -\int_{-\infty}^\infty
du (\partial_u \phis) G_{\ep}(\phis).
\label{omega4} 
\end{eqnarray}
$\Psi_2$ is immediately obtained  as
\begin{eqnarray}
\Psi_2 &=&  \int_{\phi_2}^{\phi_1} d \phi \Delta' (\phi) \nm 
         &=&  \frac{T}{2}-\frac{3}{4}T^{4/3}+O(T^{5/3}).
\label{secondterm}
\end{eqnarray}
$\Psi_4$ is also  calculated as
\begin{eqnarray}
\Psi_4 &=& 
\epsilon^2 \left. \frac{1}{4}\phi^2 
\right|_{\phi_1}^{\phi_2}
+\left. \frac{\epsilon }{2}
\left( 
\phi^2\lr{\phi-1}^2
\right)
\right|_{\phi_1}^{\phi_2} \nonumber \\
&= & \epsilon^2 \frac{1}{4}+ \frac{1}{2} \epsilon T^{2/3}.
\label{omega4}
\end{eqnarray}
Because the calculation steps for $\Psi_1$ and $\Psi_3$ are much longer 
than $\Psi_2$ and $\Psi_4$, we shall present them 
in the subsequent sections.

\subsubsection{Calculation of $\Psi_1$}          %
\label{cal_psi1}

We calculate $\Psi_1$ defined in (\ref{omega1}).
First, $\lr{\rho^{(1/2)}}^2$ in the right-hand side of (\ref{omega1}) 
is replaced with $\bra \lr{\rho^{(1/2)}}^2 \ket_{\rm \theta}$,
because ${\rho^{(1/2)}}$ is determined by the linear Langevin equation
(\ref{psiveol}).
By performing the partial integration and  using  (\ref{def_L}),
the result is expressed as
\begin{eqnarray}
\Psi_{1}
&=&
-\left.
\frac{1}{2} F'_u(\phis)\bra\lr{\rho^{(1/2)}}^2 \ket_{\rm \theta}
\right|_{u=-\infty}^{u=\infty}
\nm
& & 
 \left. 
+\frac{1}{4}\partial_u^2\bra (\rho^{(1/2)})^2\ket_{\rm \theta}
\right|_{u=-\infty}^{u=\infty} \nm
& & 
-\frac{1}{2}\int_{-\infty}^\infty  du
\hat L \partial_u \bra(\rho^{(1/2)})^2\ket_{\rm \theta}.
\label{first_term}
\end{eqnarray}
The third term of the right-hand side of (\ref{first_term}) 
is further rewritten as 
\begin{eqnarray}
& &
-\frac{1}{2}\int_{\Lambda} d\lambda \frac{T}{\lambda}
\int_{-\infty}^\infty  du
\lr{\frac{1}{2}\partial_u^2-F'_u(\phis)} \nm
& &
\times 
\lr{\Phi_\lambda(u)\partial_u\Phi_\lambda(u)^*
+\Phi_\lambda^*(u)\partial_u\Phi_\lambda(u)}
\nm
&=&
-\frac{1}{2}\int_{\Lambda}d \lambda \frac{T}{\lambda}
\left[
-\lambda|\Phi_\lambda(u)|^2
+\frac{1}{2}|\partial_u \Phi_\lambda(u)|^2 \right. \nm
& & 
\left.
\left. 
+\frac{1}{2}
\lr{
\Phi_\lambda(u)\partial_u^2\Phi_\lambda(u)^*
+\Phi_\lambda^*(u)\partial_u^2\Phi_\lambda(u)}
\right]
\right|_{u=-\infty}^{u=\infty}.
\label{first_term_3rd}
\end{eqnarray}
By substituting (\ref{first_term_3rd}) into 
the third term of (\ref{first_term})
and using the eigenvalue equation again, we express
$\Psi_{1}$ in terms of the difference of boundary 
values of a quantity $H$. That is, we write
\begin{eqnarray}
\Psi_{1}=H(\infty)-H(-\infty),
\end{eqnarray}
where $H(u)$ is defined  as 
\begin{eqnarray}
H(u)&=&
-\frac{1}{2}\int_{\Lambda}d\lambda \frac{T}{\lambda}\nm
&&
\left[
-\frac{1}{2}|\partial_u\Phi_\lambda(u)|^2
+\frac{1}{2}\Phi_\lambda^*(u)\partial_u^2\Phi_\lambda(u)
\right].
\label{Hdef}
\end{eqnarray}


Next, we calculate $H(\pm \infty)$. In terms of the Green 
function given by (\ref{Green}), we express (\ref{Hdef}) as 
\begin{eqnarray}
H(u) &=&
\frac{T}{4}
\left. \left[
\partial_z^2G(z,y)
-\partial_z\partial_yG(z,y)
\right] \right|_{z=y=u}. 
\end{eqnarray}
Here, it is reasonable to assume that 
\begin{eqnarray}
H(\pm \infty) &=&
\frac{T}{4}
\left. \left[
\partial_z^2G_{\pm}(z,y)
-\partial_z\partial_yG_{\pm}(z,y)
\right] \right|_{z=y=\pm \infty}. \nm
\end{eqnarray}
We then obtain 
\begin{eqnarray}
\Psi_{1}
&=& H(\infty)-H(-\infty) \nm
&=& \frac{T}{4}\left(-2m_++2m_-\right) \nm
&=&-\frac{T}{2}\lr{1-\sqrt{6}T^{1/3}+O(T^{2/3})}.
\label{omega1-result}
\end{eqnarray}

\subsubsection{Calculation of $\Psi_3$} \label{calculation of Psi3}

We calculate $\Psi_3$ defined in (\ref{omega3}). First, 
note that the integral region in (\ref{omega3}) is written 
in a formal manner. More precisely,
since the $u$-integration 
should be defined in the bulk region of the special solution
$\phis (u)$, the integral region is replaced with $[-\tau,\infty]$, 
where $-\tau$ corresponds to a matching point between 
the solutions $\phiB (u+\theta)$ and $\phis (u)$. Since the behaviors  of $\phiB(u+\theta)$
and $\phis(u)$ are symmetric around $\phi \simeq 1$, we assume 
that the matching point is $ \theta/2$ when $\theta \gg 1$. 
From this consideration, we set 
\begin{equation}
\tau=\frac{\theta}{2}.
\end{equation}
Note that $\tau$ becomes larger as the typical value of $\theta $ 
is larger. Indeed, $\tau \to \infty$ in the limit $T \to 0$.
The integral region $[-\infty,\infty]$ in (\ref{omega3}) should be 
read as 
a formal expression 
for $[-\tau,\infty]$ with the limit  $T \to 0$.

Now, by performing the partial integration and by using the 
relation (\ref{zero}), we obtain
\begin{eqnarray}
\Psi_3 &=&
\left. \frac{1}{2}
\partial_u\phis(u)\cdot\partial_u (\phiB(u+\theta)-\phi_2)
\right\vert_{u=-\tau}^{u=\infty}  \nm
& &
\left. -\frac{1}{2}
\lr{\partial_u^2\phis(u)} (\phiB(u+\theta)-\phi_2)
\right\vert_{u=-\tau}^{u=\infty}.
\label{omega3_1}
\end{eqnarray}
Since  $\lim_{u\to\infty} \partial_u (\phiB-\phi_2)=0$
and $\lim_{u\to\infty}  (\phiB-\phi_2)=0$, we write
\begin{eqnarray}
\Psi_3 = -\frac{1}{2}(K_1-K_2),
\end{eqnarray}
with
\begin{eqnarray}
K_1 &=& \left. 
(\partial_u \phis) \partial_u (\phiB-\phi_2) \right|_{u=-\tau},  \\
K_2 &=& \left.  
[\partial_u (\partial_u \phis) ]  (\phiB-\phi_2) \right|_{u=-\tau} .
\end{eqnarray}

Let us evaluate $K_1$ and $K_2$. First, since  the $T$ 
dependence in this contribution is not singular, we assume 
$T \to 0$ in this evaluation. Then, $\phiB$ and $\phis$ satisfy 
$\partial_t \phi=-\phi (\phi-1)^2$. 
We then derive the asymptotic form 
\begin{eqnarray}
\phiB(t) \simeq \phi_2+\frac{1}{t}
\end{eqnarray}
as $t \to \infty$, and 
\begin{eqnarray}
\phis(u) \simeq \phi_2+\frac{1}{u}
\end{eqnarray}
as $u \to -\infty$. By
using these asymptotic forms, we calculate 
\begin{eqnarray}
K_1 &=& \frac{16}{\theta^4},\\
K_2 &=& -\frac{32}{\theta^4}.
\end{eqnarray}
We thus obtain 
\begin{eqnarray}
\Psi_3 = - \frac{24}{\theta^4}.
\label{omega3-result}
\end{eqnarray}

\subsubsection{Result of $\Omega_1$} \label{The next  order result} %

\begin{figure}[htbp]
\includegraphics[width=7cm]{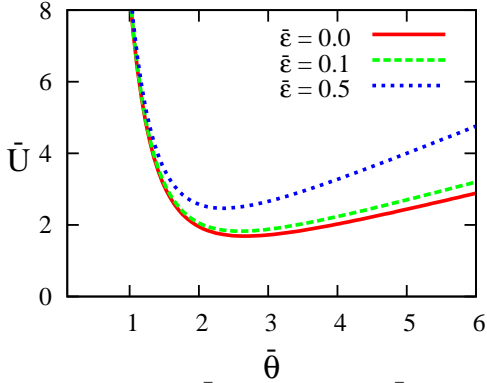}
\caption{(color online).
$\bar U$ as functions of $\bar \theta$ for 
a few values of $\bar \ep$.
}
\label{fig_pottheta}
\end{figure}

By substituting (\ref{secondterm}), (\ref{omega4}),
(\ref{omega1-result}), and (\ref{omega3-result}) into (\ref{sol:1}),  
we obtain the result of $\Omega_1$ as
\begin{equation}
-\Gamma \Omega_1
= 
\epsilon^2 \frac{1}{4}
+\frac{1}{2}\epsilon T^{2/3}
+\frac{2\sqrt{6}-3}{4}T^{4/3}
- \frac{24}{\theta^4}.
\label{sol:11}
\end{equation}
We define a potential $U(\theta; T,\epsilon)$
so that (\ref{sol:11}) is expressed by
\begin{equation}
\Gamma \Omega_1
= -\partial_{\theta} U,
\label{sol:12}
\end{equation}
where the potential is determined as 
\begin{eqnarray}
U(\theta; T,\epsilon)= 
\epsilon^2 \frac{1}{4}\theta 
+ \frac{1}{2}\epsilon T^{2/3}\theta
+\frac{2\sqrt{6}-3}{4}T^{4/3}\theta
 +\frac{8}{\theta^3}.\nm
\label{pottheta}
\end{eqnarray}
It is worthwhile noting that 
$U(\theta; T,\epsilon)$ satisfies the scaling relation
\begin{eqnarray}
U(\theta; T,\epsilon)= TU(\theta T^{1/3};1,\ep T^{-2/3}).
\label{pottheta2}
\end{eqnarray}
We then define ${\bar \theta} = \theta T^{1/3}$, 
${\bar \ep} = \ep T^{-2/3}$, and ${\bar U}(\bar \theta ;\bar \ep) 
=U(\bar \theta;1, \bar \epsilon)$. 
In Fig.~\ref{fig_pottheta}, we plot $\bar U(\bar \theta; \bar \epsilon)$
as functions of $\bar \theta$ for a few values of $\bar \epsilon$.
Here, the first, second, and third terms of  (\ref{pottheta}) 
represent the driving force in the negative direction of $\theta$,
while the last term of (\ref{pottheta}) is the repulsion from
the boundary $\theta=0$. The most probable value of $\bar \theta$,
which corresponds to the minimum of the potential, is determined 
by the balance of these two effects.

\subsection{Distribution function of $\theta$} \label{results} %

We combine (\ref{sol:1-2}) and (\ref{sol:11}).
After setting $\mu=1$, we obtain
\begin{equation}
\Gamma \partial_s \theta= 
-\epsilon^2 \frac{1}{4}
-\frac{1}{2}\epsilon T^{2/3}
-\frac{2\sqrt{6}-3}{4}T^{4/3}
+ \frac{24}{\theta^4}
+\sqrt{2 \Gamma T} \eta_\theta,
\label{final_theta}
\end{equation}
where $\eta_\theta$ satisfies (\ref{noise-theta-int}).
Using the potential (\ref{pottheta}), we derive the $s$-stationary 
distribution function of $\theta$ as 
\begin{eqnarray}
P(\theta;T,\epsilon)
= \frac{1}{Z_\theta}\exp\left(-\frac{U(\theta; T, \epsilon)}{T}\right),
\label{dist}
\end{eqnarray}
where $Z_\theta$ is the normalization constant. 
It should be noted that the distribution function satisfies
the scaling relation
\begin{eqnarray}
P(\theta;T,\epsilon)=\bar P(\theta T^{1/3};\epsilon T^{-2/3}).
\label{dist_scale}
\end{eqnarray}
The expression
(\ref{dist})  with (\ref{pottheta}) is 
the main result of our perturbative calculation.

Unfortunately,
the distribution function (\ref{dist}) with (\ref{pottheta}) 
is {\it not} the precise expression  even in the
limit $\ep \ll T^{2/3} \ll 1$.
There is a subtle
reason.  Recall that (\ref{final_theta}) is valid up to
$O(\mu)$ in the formal expansion series. When we calculate
higher order contributions in the equation for $\theta$,
the coefficients in (\ref{final_theta}) are modified. 
For instance, 
$-\Gamma \Omega_2$ includes the contribution
\begin{equation}
I=-\frac{1}{2}\int_{-\infty}^\infty  du
(\partial_u\phis(u)) F''_u(\phis)\bra (\rho^{(1)})^2\ket_{\rm \theta}.
\label{I:cont}
\end{equation}
By estimating this quantity, we have found that this 
term is $O(T^{4/3})+O(\ep^2)+O(\ep T^{2/3})$. (See Appendix \ref{hot}.)
The determination of the coefficients of these terms of $I$
is impossible without the numerical integration. Therefore, we 
did not derive the precise expression in 
the limit $\ep \ll T^{2/3} \ll 1$.
Nevertheless, we present two positive remarks. 
First, terms of $O(T^{4/3})+O(\ep^2)+O(\ep T^{2/3})$
do not appear beyond some order of expansion in $\mu$. 
Therefore, in principle,  one may have the formula determining 
the coefficients of these terms.
Second, the scaling relation 
(\ref{dist_scale}) in  the limit $\ep \ll T^{2/3} \ll 1$
seems valid up to all orders of expansion in $\mu$.

\begin{figure}[htbp]
\includegraphics[width=7cm]{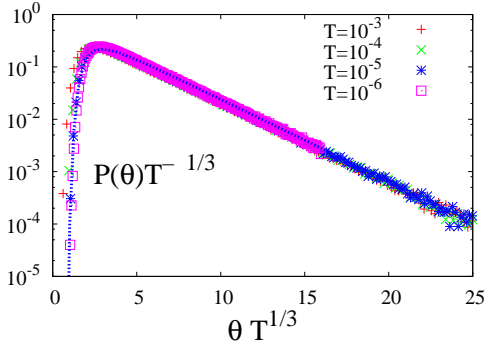}
\caption{(color online).
$
P(\theta;T,0) T^{-1/3}$ as functions of 
$\theta T^{1/3}$ for several  values of $T$.
The dashed blue line represents the functional form (\ref{guide_line})
with $a=0.36$ and $b=9.0$.
}
\label{scaling_Ptheta}
\end{figure}

In order to check the validity of the scaling relation 
(\ref{dist_scale}), we investigate the 
distribution function $P(\theta;T,\epsilon)$ for the case $\epsilon=0$
by numerical simulations of (\ref{langevin}). 
In Fig.~\ref{scaling_Ptheta}, we plotted 
$P(\theta;T,0) T^{-1/3}$ as functions 
of $\theta T^{1/3}$ for several 
values of $T$.
One may find that four graphs for different values of $T$
are not completely collapsed on one curve
in the region for small $\theta T^{1/3}$.
However, since the two curves with
$T=10^{-5}$ and $T=10^{-6}$ almost coincide with each other,
we expect that one universal curve is obtained when $T$ is decreased further.
We thus conjecture the scaling relation 
(\ref{dist_scale}) in the limit $T\to 0$ is valid.

Furthermore, following our theory, we attempt to fit
these numerical data by assuming the form
\begin{equation}
P(\theta;T,0)=\exp\left( -\frac{1}{T}\left[
 a T^{4/3}\theta
 +\frac{b}{\theta^3}\right] +c \right),
\label{guide_line}
\end{equation}
where $c$ is determined by the normalization condition
when values of $a$ and $b$ are given.
These values 
well-fitted to the numerical data are estimated as 
$a_{\rm fit}=0.36$ and $b_{\rm fit}=9$, which
are compared with our theoretical values $a_*=(2\sqrt{6}-3)/4=0.4747\cdots$
and $b_*=8$. The slight difference between $(a_{\rm fit}, b_{\rm fit})$ 
and $(a_*,b_*)$  comes  from the contribution of  higher order terms
than  $O(\mu)$. (As an example, the numerical estimation 
of $I$ in (\ref{I:cont}) provides $I \simeq 0.1 T^{4/3}$, which
might improve the difference between $a_*$ and $a_{\rm fit}$.)

\subsection{Results for the regime $T^{2/3} \ll  \ep \ll 1$} \label{remark}

The theoretical argument developed for the regime
$\ep \ll T^{2/3} \ll 1$ cannot be applied
to the regime $T^{2/3} \ll \ep \ll 1$, 
because a deviation  $\rho^{}(u) \simeq O(\ep T^{-1/3})$
becomes quite large there. (See (\ref{rho1est}) in Appendix \ref{hot}.)
In particular, from a fact 
that the unperturbed potential $\Vu(\phi)$ goes back to the 
original potential $\tilde V_{T=0}$ in  the extreme case $T=0$,
it is obvious that we need to reformulate a perturbation theory.

\begin{figure}[htbp]
\includegraphics[width=7cm]{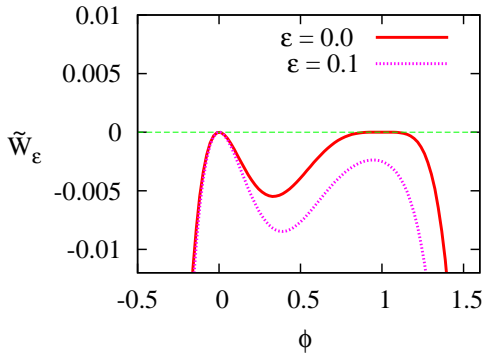}
\caption{(color online).
$\tilde W_\ep$ as functions of $\phi$.
They possess two maxima at $\phi=0$ and $\phi=1-\ep/2$.
}
\label{fig_potd2}
\end{figure}

The idea is natural and simple: 
we replace the unperturbative potential $\Vu(\phi)$
with  a potential
$\Wu(\phi)$ which is appropriate in this regime. Concretely,
instead of (\ref{ft_def}) and (\ref{vt_def}), 
we define $\tilde W_\ep (\phi)$ by 
\begin{eqnarray}
F_{\ep, T=0}(\phi)= -\partial_{\phi}\tilde W_\ep(\phi),
\label{fep_def}
\end{eqnarray}
where the functional form of $\tilde W_\ep(\phi)$
is given by
\begin{eqnarray}
\tilde W_\ep(\phi)=-\frac{1}{4}\phi^2(\phi-1)^4
-\frac{\phi^2}{4}\ep^2
-\frac{\phi^2}{2}\lr{\phi-1}^2\ep.
\label{vtep_def}
\end{eqnarray}
 Here, $\tilde W_\ep(\phi)$ has two maximum points
as displayed in Fig.~\ref{fig_potd2}.
Then, in the manner similar to  (\ref{decom}),  
we consider the decomposition
\begin{eqnarray}
\tilde W_\ep(\phi)=\Wu(\phi)+\alpha(\phi),
\label{delta_ep_def}
\end{eqnarray}
where $\alpha(\phi)$ is chosen such that 
the two maximal values  of the potential $W_{\rm u}(\phi)$ 
are identical. With this setting up, we study
the equation
\begin{eqnarray}
\partial_s\phi&=&\frac{1}{2}\partial_t^2{\phi}+\partial_\phi W_{\rm u}(\phi)
-J_{T}(\phi)+\alpha_\ep' (\phi)+\sqrt{2T}\eta.\nm
\label{langevin_s_ep}
\end{eqnarray}

After that, we repeat essentially the same procedures
under the assumption that the last three terms are treated as perturbations.
Note that the special solutions
$\phis$ and $\phi_{\rm B}$ are redefined using
$W_{\rm u}(\phi)$ instead of $V_{\rm u}(\phi)$.
Then,  in the regime $T^{2/3} \ll \ep \ll 1$,
we derive the leading order expression
\begin{eqnarray}
\Gamma \partial_s \theta = 
-\epsilon^2 \frac{1}{4}
+ \frac{24}{\theta^4}
+\sqrt{2 \Gamma T} \eta_\theta.
\end{eqnarray}
Here, the complicated terms containing $\rho$ that appeared  
in the analysis in the regime $\ep \ll T^{2/3} \ll 1$
become higher order terms in the regime $T^{2/3} \ll \ep \ll 1$.
Then, the $s$-stationary distribution function of $\theta$ is 
\begin{eqnarray}
P( \theta;T,\epsilon)= \frac{1}{Y}
\exp\left(- \frac{1}{T}\lr{ \epsilon^2 \frac{1}{4}\theta 
+\frac{8}{ \theta^3}} \right),
\label{P_dist_ep}
\end{eqnarray}
where $Y$ is the normalization constant. Setting 
$\bar \theta=\ep^{1/2} \theta$ and $\bar \epsilon =\epsilon T^{-2/3}$,
we rewrite  
$P( \theta;T,\epsilon)$ as $\bar P( \bar \theta;\bar \epsilon)$, 
where 
\begin{eqnarray}
\bar  P(\bar \theta; \bar \epsilon)= 
\frac{1}{Y}
\exp\left(- {\bar \epsilon}^{3/2} 
\left[\frac{1}{4} \bar \theta  +\frac{8}{\bar\theta^3} \right]
\right).
\label{P_dist_ep_scale}
\end{eqnarray}
Since $\bar \ep \ll 1$, 
$\bar  P(\bar \theta; \bar \epsilon)$ is further simplified to 
\begin{eqnarray}
\bar  P(\bar \theta; \bar \epsilon)= 
\frac{1}{Y}
\exp\left(- {\bar \epsilon}^{3/2} 
\lr{\bar \theta -2 (6^{1/4})}^2
\frac{1}{4(6^{1/4})}
\right).
\label{P_dist_ep_scale}
\end{eqnarray}
This is equivalent to (\ref{prob_thetas}),
where 
$\bra\theta\ket$ and $\chi_{\rm \theta}$ are determined as 
\begin{eqnarray}
\bra \theta \ket  &=& 2 (6^{1/4}) \epsilon^{-1/2} 
\label{dist_scale1}, \\
\chi_\theta &=& 2 (6^{1/4})\epsilon^{-5/2}. 
\label{dist_scale2}
\end{eqnarray}
We expect that these expressions are exact in the limit
$T^{2/3}\ll \ep \ll 1$.
In Figs. \ref{theta_scale1} and \ref{theta_scale2}, we display
these theoretical results. 
They are in 
good agreement  with the results of numerical simulations. 

\section{Concluding remarks} \label{discussion}


In this paper, we have developed a theoretical framework
for calculating statistical properties of critical 
fluctuations near a saddle-node bifurcation. The essential 
idea in our formulation is to choose an unperturbative state
in accordance with the bifurcation structure.
Since trajectories kinked in the time direction have much 
statistical weight near the bifurcation point, we express 
trajectories as
$\phi(t)=\phis(t-\theta)+(\phiB(t)-\phi_2)+\rho(t)$,
where $\phis(t-\theta)$ represents a one-parameter family of 
classical solutions in the language of the 
path-integral expression. The parameter $\theta$ is 
regarded as a Goldstone mode associated with the time-translational symmetry.
This expression naturally provides
a divergent behavior, because the Goldstone mode is
gapless or massless. Indeed, we have found that 
the fluctuation intensity of $\theta$ exhibits the 
divergence in the limit $\ep \to 0$ with small $T$ fixed 
and in the limit $T \to 0$ with $\epsilon =0$ 
fixed. The divergent behavior of $\chi_\phi(t)$ originates
from critical fluctuations of $\theta$, and $\chi_\phi(t)$ becomes
complicated due to a non-linear transformation to $\phi$
from $\theta$, as shown in Figs.~\ref{chi0} and \ref{theta_scale2}.


Before ending the paper, we wish to explain how the results
in this paper are related to understandings of other apparently 
different systems. 
First, our results suggest the following general story.
When a 
saddle in a deterministic description becomes to be connected to 
an absorbing point  at some parameter value, fluctuations of 
trajectories exhibit a critical divergence
due to the existence of a Goldstone mode; 
and if the saddle is 
far from an absorbing point, the final value of the trajectory 
exhibits a discontinuity.
Such cases might be 
related to a mixed order transition
\cite{chi4_allstar1,chi4_allstar2,silbert_mix,Toninelli_b,toninelli,
schwartz,sellitto}.
The elementary saddle-node
bifurcation studied in this paper corresponds to the simplest 
one among them. As other types of bifurcation associated with 
a mixed order transition, we list up 
a saddle-connection bifurcation which arises in a model for a many-body 
colloidal system \cite{jstat, epl}, 
and a mode-coupling transition in a spherical 
$p$-spin glass model \cite{pspin_crisanti,cugliandolo}. 
Here, the important message of our paper is that one will be able 
to develop a calculation method for the statistical 
properties of critical fluctuations at a mixed order 
transition by applying the basic idea of our formulation
to each system under investigation. 


As an interesting and non-trivial, but still simple example
of mixed order transition, 
we briefly discuss the mode-coupling 
transition of a spherical $p$-spin glass model with $p \geq 3$. 
The fluctuation property was studied by a  mode-coupling
equation supplemented with an external field
\cite{franzparisi,chi4_MCT} or by the field theoretical 
analysis \cite{chi4_BB}. Differently from these previous
methods, we will be able to consider another theoretical framework 
for critical fluctuations along with the above-mentioned 
strategy. Concretely, we start with a useful expression
of trajectories as we did.
In the thermodynamic limit,
the equation for the time-correlation function obeys a 
mode-coupling equation \cite{pspin_crisanti,cugliandolo}, and recently we have
derived a global expression of the solution near the
mode-coupling transition \cite{pspin},
which is similar to (\ref{phit_assume_model}).  
From this
result, we may express a fluctuating correlator in terms 
of a Goldstone mode $\lambda$ associated with the dilation 
symmetry that arises in the slowest time scale. Thus,
it is natural to describe critical fluctuations near
the mode-coupling transition in terms of fluctuations
of $\lambda$. The concrete calculation will be reported
in future. 


The results in this paper also provide some physical insights
into the dynamics associated with a mixed order transition 
even if the calculation has never been performed yet. For example, let us
consider a dynamical behavior of a super-cooled liquid
\cite{cavagna}.
Within a framework of the mode-coupling theory, a mode-coupling 
transition occurs at some temperature or density \cite{Gotze}. In some 
super-cooled liquids, 
the behavior associated with the transition 
has been observed approximately \cite{kob,megen}.
The physical picture is well-understood: when the  system is near 
the mode-coupling transition point, a particle cannot move freely; 
and only when the particle gets over surrounding particles, it
can move. Such an  event is called 
an {\it unlocking event} \cite{chi4_BB} or 
a {\it bond breakage event} \cite{ryamamoto}.
As the mode-coupling transition is approached,
the frequency of unlocking 
events becomes smaller and  the events become correlated more 
and more in a spatially heterogeneous manner,
which is called the {\it dynamical heterogeneity}. 
(See Refs.~\cite{chi4_allstar1,chi4_allstar2} as reviews, and
Refs.~\cite{ediger,durian,shear,dh_re_berthier}
as experimental studies, and
Refs.~\cite{ryamamoto,dh_harrowell,dh_parisi,early1,
dh_glotzer,dh_glotzer2,dh_berthier,chi4_lacevic,silbert_mix, sellitto} as numerical observations, and 
Refs.~\cite{schwartz,franzparisi,chi4_theory1,chi4_BB,chi4_MCT,garrahan,WBG,Toninelli_b,toninelli,theory_review}
as theoretical studies.)


From our point of view, we conjecture that an unlocking event near 
the mode-coupling transition might correspond to an exiting event from a marginal 
saddle in some local dynamics. Then, the results of our paper suggest
that the most important characterization of the dynamical heterogeneity
is space-time fluctuations of exiting time from a saddle-point. 
At present, it seems difficult to derive such local dynamics 
theoretically, but it is tempting to connect the idea of 
cooperative arrangement regions to 
a marginal saddle in some local dynamics. Since the distribution 
function of exiting time from cooperative arrangement regions exhibits
an interesting behavior \cite{matsui},
the theoretical derivation of the observation may be a good starting
point for the consideration. 
The final goal in this direction is to obtain a simple
expression determining space-time fluctuations of the
Goldstone mode on the basis of a microscopic particle
model. We will study it step by step toward this goal.


Related to the description of dynamical heterogeneity, we remark on
a well-recognized conjecture that the mode-coupling transition 
described theoretically is nothing but 
a cross-over phenomenon in finite dimensional systems. In order to
understand the nature of this cross-over, one needs to describe 
an activation process from ``pseudo" meta-stable states 
which are not defined clearly, but might be connected to that 
defined in the mean-field approximation.
Although, physically, the activation process  corresponds to a 
nucleation of some domain, its mathematical expression is highly 
non-trivial. Here, 
when we apply our analysis to a finite dimensional system, 
in our viewpoint, the cross-over phenomenon
is equivalent to the finite value of the expectation of the Goldstone 
mode. Thus, we have only to calculate an asymptotic tail of the effective 
potential for the Goldstone mode in the limit $t \to \infty$. Since the analysis 
of the super-cooled liquid is too difficult, we should begin with the study
of  a diffusively coupled model of local dynamics (\ref{langevin}) 
with (\ref{fformm}). (See Ref. \cite{1dimlett} as a report on 
a numerical experiment with $\ep=0$.) It would be possible to
develop the mean-field analysis of the spatially extended system, but
it seems difficult  to treat spatial fluctuations accurately even for 
such a simple system. To develop a systematic theory beyond the mean
field analysis is a challenging problem.


Finally, let us recall that our formulation is based on
the fictitious time formalism. One may expect that 
the calculation can be done within a standard Martin-Siggia-Rose (MSR)
formalism \cite{msr}. Such a reformulation is particularly important
when we study more complicated systems. In this context,
it is worthwhile noting that the third term in (\ref{pottheta}) 
has been obtained in the MSR formulation with a semi-classical 
approximation \cite{fukui}.
Such calculation techniques 
in spatially extended systems will be developed in future.


\begin{acknowledgments} 
The authors acknowledge T. Fukui and K. Takeuchi for useful communications.
This work was supported by a grant from 
the Ministry of Education, Science, Sports and Culture of Japan, 
Nos. 19540394
and 21015005. Mami Iwata acknowledges the support by Hayashi 
memorial foundation for female natural scientists.
\end{acknowledgments}

\appendix

\section{Path integral expression} \label{path}

We derive the path-integral expression (\ref{path_integral1}) 
with (\ref{path_integral2}) from the Langevin equation (\ref{langevin})
with (\ref{gauss_noise_model}). In particular, we carefully discuss 
the derivation of the so-called Jacobian term. 

Let $\Delta t$ be a sufficiently small time interval. 
We discretize (\ref{langevin}) as 
\begin{eqnarray}
\phi_{n}-\phi_{n-1}
&=&[f_\ep(\phi_{n-1})+f_\ep(\phi_n)]\frac{\Dt}{2}\nm
&&+\xi_{n-1}\Dt+O((\Dt)^2),
\label{model:dis1}
\end{eqnarray}
with $n=1,2,\cdots, N$.
Here, $\lr{\xi_n}_{n=0}^{N-1}$ obeys
the Gaussian distribution 
\begin{eqnarray}
P_\xi(\xi_0,\xi_1,\cdots,\xi_{N-1}) 
= \left(\frac{\Dt}{4\pi T} \right)^{N/2}
\exp\left(-\frac{\Dt}{4T}\sum_{n=0}^{N-1} \xi_n^2\right). \nm
\end{eqnarray}
In the limit $\Delta t \to 0$, $\phi_n$ is expected to 
provide $\phi(n \Delta t)$ in the Langevin equation.

Let us fix $\phi_0$. Then, a sequence $(\xi_0,\xi_1,\cdots,\xi_{N-1})$
determines uniquely the sequence $(\phi_1,\cdots,\phi_N)$.  
Thus, the distribution function of the sequence 
$(\phi_n)_{n=1}^N$ is expressed as 
\begin{eqnarray}
P(\phi_1,\cdots,\phi_{N}) =P_\xi(\xi_0,\cdots,\xi_{N-1}) 
\left\vert \pder{(\xi_0,\cdots,\xi_{N-1})}{(\phi_1,\cdots, \phi_N)} 
\right\vert.\nm
\end{eqnarray}
Here, the determinant of the Jacobian matrix is calculated as
\begin{equation}
\left
\vert \pder{(\xi_0,\cdots,\xi_{N-1})}{(\phi_1,\cdots, \phi_N)} 
\right\vert
=  \prod_{n=1}^N 
\left\vert 1-\frac{1}{2}f_\ep'(\phi_n) 
\Dt \right\vert \frac{1}{(\Dt)^N} .
\end{equation}
By using the relation
\begin{eqnarray}
&&\prod_{n=1}^N 
\left\vert 1-\frac{1}{2}f_\ep'(\phi_n) 
\Dt \right\vert \nm
&=&   1- \sum_{n=1}^N \frac{1}{2} f_\ep'(\phi_n) \Dt  +O((\Dt)^2) 
 \nonumber \\
&=& \exp\left(- \sum_{n=1}^N \frac{1}{2} f_\ep'(\phi_n) \Dt
+O( (\Dt)^2 ) \right) , 
\end{eqnarray}
we obtain
\begin{eqnarray}
& & P(\phi_1,\cdots,\phi_{N}) =
\left(\frac{1}{4\pi T\Dt} \right)^{N/2} \nm
&&\exp
\left(
       -\frac{\Dt}{4T}
        \sum_{n=1}^{N} 
\left\{
\right. 
\right. 
\nm
&&
\left.
\left[
\frac{\phi_{n}-\phi_{n-1}}{\Dt}-(f_\ep(\phi_{n-1})+f_\ep(\phi_n))/2 
\right]^2 
\right. 
\nonumber \\
& & 
\left.
\left.
+2T  f_\ep'(\phi_n) +O((\Dt)^{1/2}) 
\right\}
\right). 
\label{path_6}
\end{eqnarray}
By taking the limit $\Dt \to 0$ and $N \to \infty$
with $N\Dt $ fix, we write 
formally (\ref{path_integral1}) with (\ref{path_integral2}). 

At the end of this appendix, we remark on a discretization method.
One may notice that another discretized expression
\begin{eqnarray}
\phi_{n}-\phi_{n-1}=f_\ep(\phi_{n-1})\Dt +\xi_{n-1}\Dt +O( (\Dt)^{3/2})
\label{model:dis2}
\end{eqnarray}
does not yield the Jacobian term, because in this case the determinant
of the Jacobian matrix  
\begin{eqnarray}
\left\vert 
\pder{(\xi_0,\cdots,\xi_{N-1})}{(\phi_1,\cdots,\phi_N)} 
\right\vert
=\frac{1}{(\Dt)^N}
\end{eqnarray}
does not depend on $\phi$. However, the discretization 
 (\ref{model:dis2}) provides
\begin{eqnarray}
& & P(\phi_1,\cdots,\phi_{N}) = \frac{1}{(\Dt)^N} 
\left(\frac{\Dt}{4\pi T} \right)^{N/2}  \nonumber \\
& & \exp
\left(
       -\frac{\Dt}{4T}
        \sum_{n=1}^{N} 
\right. 
\nm
&&
\left.
\left\{
\left[
\frac{\phi_{n}-\phi_{n-1}}{\Dt}-f_\ep(\phi_{n-1}) 
\right]^2 +R
\right\}
\right),
\label{final2} 
\end{eqnarray}
where  the term $R$ comes from the product of $\phi_n-\phi_{n-1}$ 
and the last term $O((\Delta t )^{3/2})$ in the right-hand side
of (A7) when $\xi_{n-1}^2$ is evaluated. Explicitly, $R$
is equal to $O( (\Dt)^{3/2})(\phi_{n}-\phi_{n-1})/(\Dt)^2$.
We here note that $\phi_{n}-\phi_{n-1} \simeq O((\Delta t)^{1/2})$.
This leads to $R =O( (\Delta t)^0)=O(1)$ in the limit $\Delta t\to 0$.
Therefore, (\ref{final2}) is not useful in the limit $\Delta t  \to 0$.
With regard to the discretization problem, we remark that
numerical simulations of the discretized 
form (\ref{model:dis2}) yield (\ref{scaling_T3}) 
and (\ref{scaling_T4}), too. 
Here, one may confirm that (\ref{scaling_T3}) and (\ref{scaling_T4})
cannot be obtained without the Jacobian term in (\ref{path_integral2}) 
within the analysis of the path integral expression.
Therefore, this example provides an evidence for the claim that 
the path integral expression in the limit $\Dt \to 0$
always contains the last term in   (\ref{path_integral2}).

\section{estimation of $I$} \label{hot}

We estimate the integral $I$ 
given in (\ref{I:cont}). Before the estimation,
we need to evaluate $\rho^{(1)}$. 
We multiply by $\Phi_{\lambda}^* (u)$ on both sides of 
(\ref{eq:1}) and integrate them over $-\infty \leq u\leq \infty$.
We then consider the $s$-stationary state and extract
the lowest order terms. By using (\ref{rho-exp}),
we derive
\begin{eqnarray}
\lambda \bra\psi_{\lambda}^{(1)}\ket_{\theta} &=&
-\frac{1}{2}\int_{-\infty}^\infty  du \Phi_{\lambda}^* (u)
\Fu''(\phis(u))\bra\lr{\rho^{(1/2)}}^2
\ket_{\theta}
\nm
&&
-\int_{-\infty}^\infty  du \Phi_{\lambda}^* (u)
G_{\ep}(\phis(u)),
\label{ave_rho_app_0}
\end{eqnarray}
where the contribution from the fourth and fifth terms
in the right hand side of (\ref{eq:1}) have been neglected,
because they are estimated as higher order terms.
From (\ref{rho-exp}), 
(\ref{rho_1/2_green}), and  (\ref{Green}), we obtain
\begin{eqnarray}
\bra \rho^{(1)}(u) \ket_\theta  
&=&-\frac{T}{2} \int_{-\infty}^\infty  
dv G(u,v)\Fu''(\phis(v))G(v,v) \nm
&& + \int_{-\infty}^\infty  dv G(u,v)G_{\ep}(\phis(v)).
\label{ave_rho_app_0}
\end{eqnarray}
Since the most singular behavior arises  around $u \to -\infty$,
we estimate $\bra \rho^{(1)}(u) \ket $ in the limit $u\to -\infty$
by noting (\ref{green_explititform}).
The first term in (\ref{ave_rho_app_0}) is estimated as 
$T \xi \times 1/m_- \times F''_{0}\times 1/m_- $, where $\xi$
is the length scale over which the integral is dominant. 
By using 
$\xi \sim 1/m_- \sim O(T^{-1/3})$
(see (\ref{def_m_pm})), we estimate 
the first term of (\ref{ave_rho_app_0}) as 
\begin{eqnarray}
&&-\frac{T}{2} \int_{-\infty}^\infty  
dv G(u,v)\Fu''(\phis(v))G(v,v) \nm
&\simeq&
O\lr{T\times T^{-1/3}\times T^{-1/3} \times  T^{1/3} 
\times T^{-1/3}} \nm
&\simeq& O\lr{T^{1/3} }.
\label{first}
\end{eqnarray}
Similarly, the second 
term in (\ref{ave_rho_app_0}) is estimated as
\begin{eqnarray}
&& \int_{-\infty}^\infty  dv G(u,v)G_{\ep}(\phis(v))\nm
&\simeq&
\xi \times 1/m_- \times G_{\ep}(\phis(u)) \nm
&\simeq & O\lr{T^{-1/3}\times T^{-1/3} \times \ep T^{1/3}  }\nm
&\simeq  &O\lr{\ep T^{-1/3}}.
\label{second}
\end{eqnarray}
Thus, we write 
\begin{equation}
\bra\rho^{(1)}(u \to -\infty)\ket_{\theta} 
= O(T^{1/3}) +O(\ep T^{-1/3}).
\label{rho1est}
\end{equation}

From this result, the quantity $I$ in (\ref{I:cont}) is 
estimated as follows. First, (\ref{I:cont}) is rewritten as 
\begin{eqnarray}
I &=&
-\frac{1}{2}\int_{-\phi_2}^{\phi_1} d\phis 
F''_u(\phis)\bra\rho^{(1)}(u(\phis)) \ket^2_{\theta},
\end{eqnarray}
where $u(\phis)$ is the inverse function of $\phis(u)$.
In this integration, the most dominant contribution
arises from the integral region $[-\phi_2,1]$,
and this contribution 
is estimated as 
\begin{eqnarray}
I&\simeq&
O\lr{T^{1/3} \times T^{1/3} \times (T^{1/3} +\ep T^{-1/3})^2}\nm
&\simeq& O(T^{4/3})+O(\ep T^{2/3})+O(\ep^2).
\end{eqnarray}



\begin{thebibliography}{99}


\bibitem{GL}
N. Goldenfeld, {\it Lectures on Phase Transitions and the Renormalization 
Group}, (Addison-Wesley, New York, 1992).

\bibitem{kuramoto}
 Y. Kuramoto, {\it Chemical Oscillations, Waves, and Turbulence},
(Springer, Berlin, 1984). 

\bibitem{Munos}
M. A. Munoz, {\it Advances in Condensed Matter and Statistical Physics}, 
ed. E. Korutcheva and R. Cuerno, (Nova Science Publishers, New York, 2004), 37 (2004).
\bibitem{Gucken}
J. Guckenheimer and P. Holmes, {\it Nonlinear Oscillations, 
Dynamical Systems and Bifurcations of Vector Fields}
(Springer-Verlag, New York, 1983). 


\bibitem{binder}
K. Binder, Phys. Rev. B {\bf 8}, 3423 (1973). 

\bibitem{reimann}
P. Reimann, C. Van den Broeck, H. Linke, P. H\"{a}nggi, J. M. 
Rubi, and A. Perez-Madrid, Phys. Rev. E {\bf 65}, 031104  (2002).


\bibitem{Tyson} J. J. Tyson, K. C. Chen, and B. Novak, Current Opinion in Cell Biology {\bf 15}, 221 (2003).
\bibitem{spike} B. Lindner, A. Longtin, and A. Bulsara, Neural Computation {\bf 15},  1761 (2003).
\bibitem{Ohta_co} H. Ohta  and S. Sasa, Phys. Rev. E {\bf 78}, 065101(R) (2008).
\bibitem{bio3} P. B. Warren, Phys. Rev. E {\bf 80}, 030903(R) (2009).


\bibitem{kcore} M. Iwata and S. Sasa, J. Phys. A: Math. Theor. {\bf 42}, 075005 (2009).
\bibitem{rfohta} H. Ohta and S Sasa, arXiv:0912.4790.
\bibitem{1dimlett} M. Iwata and S. Sasa, Phys. Rev. E {\bf 78}, 055202(R) (2008).

\bibitem{kubokitahara}
R. Kubo, K. Matsuo, and K. Kitahara, J. Stat. Phys. {\bf 9}, 51 (1973).




\bibitem{chi4_allstar1} 
L. Berthier, G. Biroli, J. P. Bouchaud, W. Kob, K. Miyazaki, and D. R. Reichman,
J. Chem. Phys. {\bf 126}, 184503 (2007).
\bibitem{chi4_allstar2} 
L. Berthier, G. Biroli, J. P. Bouchaud, W. Kob, K. Miyazaki, and D. R. Reichman,
J. Chem. Phys. {\bf 126}, 184504 (2007). 
\bibitem{silbert_mix} 
C. S. O'Hern, L. E. Silbert, A. J. Liu, and S. R. Nagel, Phys. Rev. E {\bf 68}, 011306 (2003). 

\bibitem{sellitto}
M. Sellitto, G. Biroli, and C. Toninelli, Europhys. Lett. {\bf 69}, 496 (2005).
\bibitem{schwartz}
J. M. Schwarz, A. J. Liu, and L. Q. Chayes, Europhys. Lett. {\bf 73}, 560 (2006).
\bibitem{Toninelli_b} 
C. Toninelli, G. Biroli, and D. S. Fisher, Phys. Rev. Lett. {\bf 96}, 035702 (2006). 
\bibitem{toninelli}
C. Toninelli and G. Biroli, J. Stat. Phys. {\bf 130}, 83 (2008).





\bibitem{heun}
A. Greiner, W. Strittmatter, and J. Honerkamp, 
J. Stat. Phys {\bf 51}, 95 (1988).



\bibitem{Gardner}
C. W. Gardiner, {\it Handbook of Stochastic Methods: for Physics, Chemistry and the Natural Sciences (Springer Series in Synergetics) 3rd. ed.}, 
(Springer, Berlin, 2004).

\bibitem{fictitious}
G. Parisi and W. Yongshi, Sci. Sin. {\bf 24}, 483 (1981).

\bibitem{ohtakawasaki}
K. Kawasaki and T. Ohta, Physica A { \bf 116}, 573 (1982).

\bibitem{eiohta} S. Ei and T. Ohta, Phys. Rev. E {\bf 50}, 4672 (1994). 




\bibitem{polymer} G. Costantini and F. Marchesoni, Phys. Rev. Lett. {\bf 87}, 114102 (2001).

\bibitem{kuramoto_p}
Y. Kuramoto, Prog. Theor. Phys. Suppl. {\bf 99}, 244 (1989).
\bibitem{hohenberg} M. C. Cross and P. C. Hohenberg, Rev. Mod. Phys. {\bf 65}, 851 (1993).








\bibitem{jstat} M. Iwata and S. Sasa, J. Stat. Mech. L10003 (2006).
\bibitem{epl} M. Iwata and S. Sasa, Europhys. Lett. {\bf 77}, 50008 (2007).

\bibitem{pspin_crisanti}
A. Crisanti, H. Horner, and H. J. Sommers, Z. Phys. B {\bf 92}, 257 (1993).
\bibitem{cugliandolo} L. F. Cugliandolo and J. Kurchan, Phys. Rev. Lett. {\bf 71}, 173 (1993).


\bibitem{franzparisi}
S. Franz and G. Parisi, J. Phys. Condens. Matter {\bf 12}, 6335 (2000).
\bibitem{chi4_MCT} G. Biroli, J. P. Bouchaud, K. Miyazaki, and D. R. Reichman, Phys. Rev. Lett. {\bf  97}, 195701 (2006). 

\bibitem{chi4_BB} G. Biroli and J. P. Bouchaud, Europhys. Lett. {\bf 67}, 21 (2004). 

\bibitem{pspin} M. Iwata and S. Sasa, J. Phys. A: Math. Theor. {\bf 42}, 245001 (2009).

\bibitem{cavagna} A. Cavagna, Physics Reports {\bf 476}, 51 (2009). 



\bibitem{Gotze}
W. G\"{o}tze, {\it Liquids, Freezing and Glass Transition}, 
ed D. Levesque et al (Elsevier, New York, 1991).


\bibitem{kob} 
W. Kob, {\it Slow Relaxations and Nonequilibrium Dynamics in Condensed Matter (Les Houches 2002 
Session LXXVII) },
ed J. L. Barrat et al (Berlin: Springer, 2003), 199 (2003). 

\bibitem{megen}
W. V. Megen and S. M. Underwood, Phys. Rev. Lett. {\bf 70}, 2766 (1993).


\bibitem{ryamamoto}
R. Yamamoto and A. Onuki, Phys. Rev. E {\bf 58}, 3515 (1998).

\bibitem{ediger} M. D. Ediger, Ann. Rev. Phys. Chem. {\bf 51}, 99 (2000). 

\bibitem{shear} 
O. Dauchot, G. Marty, and G. Biroli, Phys. Rev. Lett. {\bf 95}, 265701 (2005).

\bibitem{dh_re_berthier}
L. Berthier, G. Biroli, J. P. Bouchaud, L. Cipelletti, D. El Masri, D. L'H\^{o}te, F. Ladieu, and M. Pierno, Science {\bf 310}, 1797 (2005).

\bibitem{durian}
A. Abate and D. Durian, Phys. Rev. E {\bf 74}, 031308 (2006).






\bibitem{early1} S. Butler and P. Harrowell, J. Chem. Phys. {\bf 95}, 4454 (1991). 
\bibitem{dh_harrowell} M. Hurley and P. Harrowell, Phys. Rev. E {\bf 52}, 1694 (1995).
\bibitem{dh_parisi} G. Parisi, J. Phys. Chem. B {\bf 103}, 4128 (1999).
\bibitem{dh_glotzer}
C. Bennemann, C. Donati, J. Bashnagel, and S. C. Glotzer, Nature (London) 
{\bf 399}, 246 (1999).
\bibitem{dh_glotzer2} S. C. Glotzer, J. Non-Cryst. Solids {\bf 274}, 342 (2000).

\bibitem{chi4_lacevic} 
N. La\v{c}evi\'c, F. W. Starr, T. B. Schr\o{}der, and S. C. Glotzer, J. Chem. Phys. {\bf 119}, 7372 (2003). 
\bibitem{dh_berthier}
L. Berthier, Phys. Rev. E {\bf 69}, 020201(R) (2004).



\bibitem{chi4_theory1} 
C. Donati, S. Franz, G. Parisi, and S. C. Glotzer, J. Non-Cryst. Solids {\bf 307}, 215 (2002). 
\bibitem{WBG}
S. Whitelam, L. Berthier, and J. P. Garraham, Phys. Rev. Lett. {\bf 92}, 185705 (2004).
\bibitem{theory_review}
C. Toninelli, M. Wyart, L. Berthier, G. Biroli, and J. P. Bouchaud, Phys. Rev. E {\bf 71}, 041505 (2005).

\bibitem{garrahan}
A. C. Pan, J. P. Garrahan, and D. Chandler, Phys. Rev. E {\bf 72}, 041106 (2005). 




\bibitem{matsui}
J. Matsui, private communication.

\bibitem{msr}
P. C. Martin, E. D. Siggia, and H. A. Rose, Phys. Rev. A {\bf 8}, 423 (1973).

\bibitem{fukui}
T. Fukui, private communication.





\end{thebibliography}
\end{document}